\documentstyle[prl,aps,graphicx,cite,amssymb,epsfig,float,psfrag,multirow,amsmath,subfigure]{revtex}
\catcode`@=11

\def\@citex[#1]#2{\if@filesw\immediate\write\@auxout{\string\citation{#2}}\fi
  \def\@citea{}\@cite{\@for\@citeb:=#2\do
    {\@citea\def\@citea{,\penalty\@m}\@ifundefined
      {b@\@citeb}{{\bf ?}\@warning
       {Citation `\@citeb' on page \thepage \space undefined}}%
\hbox{\csname b@\@citeb\endcsname}}}{#1}}

\def\citer{\@ifnextchar [{\@tempswatrue\@citexr}{\@tempswafalse\@citexr[]}}

\def\@citexr[#1]#2{\if@filesw\immediate\write\@auxout{\string\citation{#2}}\fi
  \def\@citea{}\@cite{\@for\@citeb:=#2\do
    {\@citea\def\@citea{--\penalty\@m}\@ifundefined
       {b@\@citeb}{{\bf ?}\@warning
       {Citation `\@citeb' on page \thepage \space undefined}}%
\hbox{\csname b@\@citeb\endcsname}}}{#1}}
\catcode`@=12

\newcommand{\beq}{\begin{equation}}
\newcommand{\eequ}{\end{equation}}
\newcommand{\eeq}{\end{equation}}
\def\bea{\begin{eqnarray}}
\def\eea{\end{eqnarray}}
\def\Q{{\cal O}}
\def\nn{\nonumber}
\newcommand{\noi}{\noindent}

\begin{document}

\thispagestyle{empty} 

\begin{flushright}
\begin{tabular}{l}
{\tt DAPNIA-02-380}\\
{\tt LPT Orsay 02-122}\\
{\tt LMU-02-19}\\
{\tt PCCF-RI-0218}\\
{\tt hep-ph/0301165}\\
\end{tabular}
\end{flushright}
\vskip 2.2cm\par
\begin{center}
{\par\centering \textbf{\LARGE Testing QCD factorisation and
charming penguins in charmless ${\boldsymbol{B\to PV}}$ } }\\
\vskip 0.9cm\par
{\par\centering \large  
R.~Aleksan$^a$\footnote{e-mail~:aleksan@hep.saclay.cea.fr},
P.-F.~Giraud$^a$\footnote{e-mail~:giraudpf@hep.saclay.cea.fr},
V.~Mor\'enas$^b$\footnote{e-mail~:morenas@clermont.in2p3.fr}, 
O.~P\`ene$^c$\footnote{e-mail~:pene@th.u-psud.fr} and
A. S. Safir$^d$\footnote{e-mail~:safir@theorie.physik.uni-muenchen.de}.
}
{\par\centering \vskip 0.5 cm\par}
{\par\centering \textsl{ 
$^a$~CEA Saclay, DAPNIA/SPP (B\^at. 141)
F-91191 Gif-sur-Yvette CEDEX, France.
\\
$^b$~LPC, Universit\'e Blaise Pascal - CNRS/IN2P3 F-63000 Aubi{\`e}re Cedex, France. \\
$^c$~LPT (B\^at.210), Universit\'e de Paris XI, Centre d'Orsay, 91405
Orsay-Cedex, France. \\  
$^d$~LMU M\"unchen, Sektion Physik,Theresienstra\ss e 37, D-80333
M\"unchen, Germany.} \\ 
\vskip 0.3cm\par}
\today
 \end{center}

\vskip 0.45cm
\begin{abstract}
We try a global fit  of the experimental branching ratios  and CP-asymmetries
of  the charmless $B\to PV$  decays according
 to QCD factorisation. We find it impossible to reach  a satisfactory
 agreement, the confidence level (CL) of the best fit is smaller than .1 \%.
  The main reason for this  failure is the difficulty
 to accomodate several large  experimental branching ratios of the strange
 channels. Furthermore, experiment was not able to exclude a large direct CP
asymmetry in $\overline {B^0}\to\rho^+ \pi^-$, which is predicted very small
by QCD factorisation. Trying a fit with QCD factorisation 
 complemented  by a charming-penguin inspired model we reach a best fit 
 which is not excluded by experiment (CL of about 8 \%) but is not fully convincing.
  These negative results must be tempered  by the remark that some of the
  experimental data used are recent and might still  evolve significantly. 

\end{abstract}

\newpage \pagestyle{plain} 

\section{Introduction}
It is an important theoretical challenge to master the 
non-leptonic decay amplitudes and particularly $B$ 
non-leptonic decay. It is not only important {\it per se}, in view
of the many experimental branching ratios which have been 
measured recently with increasing accuracy by
BaBar\citer{0107058,BaBar}, Belle\citer{Belle,0201007} and
CLEO\citer{0101032,CLEO}, but it 
 is also necessary in order to get control over the measurement of  
 CP violating parameters and particularly the so-called
 angle $\alpha$ of the unitarity triangle. It is well known that
 extracting $\alpha$ from measured indirect CP asymmetries needs a
  sufficient control of the relative size of the so-called 
  tree ($T$) and penguins ($P$) amplitudes. 
 
However the theory of non-leptonic weak decays is   a difficult issue.
Lattice QCD gives predictions for semi-leptonic or purely leptonic
decays but not directly for non-leptonic ones. 
Since long, one has used what  is  now called ``naive factorization'' which 
replaces the matrix element of a four-fermion operator in a heavy-quark 
decay by the product of the matrix elements of two currents, one semi-leptonic
matrix element and one purely leptonic. 
For long it was noticed that  naive factorization did provide reasonable 
results although it was impossible to derive it rigorously
from QCD except in the $N_c \to  \infty$ limit. It was also well-known
that the matrix elements computed via naive factorization 
have a wrong anomalous dimension.
 
Recently an important theoretical progress has been performed~\cite{QCDF,0104110}
which is commonly called ``QCD factorisation''. It is based on the fact that 
the $b$ quark is heavy compared to the intrinsic scale of strong 
interactions. This allows  to deduce that non-leptonic decay amplitudes 
in the heavy-quark limit have a simple structure. It implies 
that corrections termed ``non-factorizable'', 
which were thought to be intractable, can be calculated rigorously. 
The anomalous dimension of the matrix elements is now correct to the order
at which the calculation is performed. 
Unluckily the subleading
 ${\cal  O}(\Lambda/m_b)$ contributions cannot in general be computed 
 rigorously because of infrared singularities,
  and some of these which are chirally enhanced
 are not small, of  order $ {\cal  O}(m_\pi^2/m_b(m_u+m_d))$, which shows 
 that the inverse $m_b$ power is compensated by $m_\pi/(m_u+m_d)$. 
 In the seminal papers~\cite{QCDF,0104110}, these contributions are simply 
 bounded according to a qualitative argument which could as well justify
 a significanlty larger bound  with the risk of seeing these unpredictable 
 terms become dominant.
  It is then  of utmost importance to check experimentally  QCD factorisation.

Since a few years it has been applied to $B\to PP$ (two charmless pseudoscalar
 mesons)
decays. The general feature is that the decay to non-strange final states 
is predicted slightly larger than experiment while the decay to strange
final states is significantly underestimated. In~\cite{0104110} it is claimed that
this can be cured by a value of the unitarity-triangle angle $\gamma$ larger
than generally expected, larger maybe than 90 degrees. Taking also into account
various uncertainties the authors conclude positively as for the agreement of 
QCD factorisation with the data. In~\cite{Isola:2001bn,Ciuchini:2001gv} it was objected that the large branching ratios
for strange channels argued in favor of the presence of a specific non
perturbative contribution called ``charming
penguins''~\citer{Ciuchini:2001gv,Ciuchini:1997hb}. We will return to
this approach later.

The  $B\to PV$ (charmless pseudoscalar + vector mesons) channels are
 more numerous and allow a more extensive check. In ref. \cite{Aleksan:1995wn} it was shown that 
naive factorisation implied a rather small 
$|P|/|T|$ ratio,  for  $\overline {B^0}\to\rho^\pm\pi^\mp$ decay 
channel, to be
compared to the larger one for the  $B\to\pi^+\pi^-$. 
This prediction is still valid in QCD factorisation where 
the $|P|/|T|$ ratio is of about 3 \% (8 \%) for the 
$\overline {B^0}\to\rho^+\pi^-$ ($\overline {B^0}\to\rho^-\pi^+$)  channel
against about 20 \% for the $\overline {B^0}\to\pi^+\pi^-$ one.
If this
prediction was reliable it would put the $\overline {B^0}\to\rho^+\pi^-$ channel
in a  good position to measure the CKM angle $\alpha$ via indirect
CP violation. This remark triggered the present work: we wanted to 
check QCD factorisation in the $B\to PV$ sector to  estimate the chances 
for a relatively easy determination of the angle $\alpha$.

 The non-charmed $B\to PV$  amplitudes have been 
computed in naive factorisation~\cite{ali}, in some extension of 
naive factorisation including strong phases~\cite{Kramer:1994in},
in QCD
factorisation~\citer{Yang:2000xn,Du:2002cf} and some of them in the
so-called perturbative
QCD~\cite{Chen:2001pr,Keum:2002qj}. In~\cite{Cottingham:2001ax}, a 
global fit to $B\to PP,PV,VV$ was investigated using QCDF in 
the heavy quark limit  and it has been found a
plausible set of soft QCD parameters that apart from three pseudoscalar
vector channels, fit well the experimental branching ratios.
Recently~\cite{Du:2002cf} it was
claimed from a global fit to $B\to PP, PV$ that the
predictions of QCD factorization are in good agreement with 
experiment when one excludes some channels from the global fit. When this paper appeared we had been for some time considering
this question and our feeling was significantly less optimistic. 
This difference shows that the matter is far from trivial mainly because 
 experimental uncertainties can still be open to some discussion.
We would like in this paper to understand better the origin of the difference
between our unpublished conclusion and the one presented in~\cite{Du:2002cf}  and
try to settle the present status of the comparison of QCD factorisation with experiment.

One general remark  about QCD  factorisation  is that it yields predictions which
do not differ so much from naive factorisation ones. This is expected since
QCD factorisation makes a perturbative expansion the zeroth order of
which being naive factorisation. As  a consequence, QCD factorisation
predicts very small direct CP violation in the non-strange channels.
Naive factorisation  predicts  vanishing direct CP violation.
Indeed, direct CP violation needs the occurence of two distinct 
strong contributions with a strong phase between them. 
It vanishes when the subdominant strong contribution vanishes and
also when the relative strong phase does as is the case in naive factorisation.
 In the case of non-strange
decays, the penguin  ($P$) and tree ($T$) contributions being at the same order in
Cabibbo angle, the penguin is  strongly suppressed because the Wilson 
coefficients are suppressed by at least  one power of the strong coupling 
constant $\alpha_s$, and the strong phase in QCD factorisation  is generated by 
a ${\cal O}(\alpha_s)$ corrections. Having  both  $P/T$ and the strong phase
small, the direct CP asymetries are doubly suppressed. Therefore a sizable
experimental direct CP asymetry in $\overline {B^0}\to \rho^+\pi^-$ which
is not excluded by experiment~\cite{Aubert:2002jq}
would be at odds with QCD factorisation.  We will discuss this later on.
Notice that this argument is independent of the value of the unitarity angle
$\gamma$, contrarily to  arguments based on the value of some branching
ratios which depend on $\gamma$~\cite{0104110}.

The Perturbative QCD (PQCD) predicts larger direct CP asymmetries than QCDF
 due to the fact that penguin contributions to anihilation diagrams, claimed to be
 calculable in PQCD, contribute 
 to a larger amount to the amplitude and have
a large strong phase. In fact, in PQCD, this penguin anihilation diagram 
is claimed to be of the same order, ${\cal O}(\alpha_s)$, than the dominant naive factorisation 
diagram  while in QCDF it is also  ${\cal O}(\alpha_s)$
but smaller than the  dominant naive factorisation which is  ${\cal O}(1)$.
Hence,  in PQCD, this large penguin contribution with  a large strong phase
yields a large CP asymmetry~\cite{Nagashima:2002ia,0004173,0004213}.

If QCD factorisation is concluded to be unable to describe 
the present data satisfactorily, while there is to our knowledge 
no theoretical argument against it, we have to incriminate 
non-perturbative contributions which are larger than expected.
One could simply enlarge the allowed bound for those  contributions which 
are formally subleading but might be large.  However a simple factor two on 
these bounds makes these unpredictable contributions comparable in size with
the predictable ones, if not larger. This spoils the predictivity of the whole
program. 

A second line is to make some model about the non-perturbative contribution.
The ``charming penguin'' approach~\cite{Ciuchini:1997rj,Ciuchini:1997hb}
starts from noticing the underestimate of strange-channels 
branching ratios by the  factorisation approaches. This will be shown to apply 
to the $PV$ channels as well as to the $PP$ ones. This has triggered us
to try a charming-penguin inspired approach.
 It is assumed that some hadronic contribution to the penguin loop
 is non-perturbative. In other words that weak interactions create
  a charm-anticharm intermediate state which turns into non-charmed final
  states by strong rescattering.   In order to make the model  as predictive as
  possible we will use not more than two unkown complex number and  use 
  flavor symmetry in strong rescattering. 
 
In section \ref{sec:Heff} we will recall the weak-interaction 
effective Hamiltonian. In section \ref{qcdeff} we will recall QCD factorisation.
In section \ref{exp} we will  compare QCD factorisation with experimental 
branching ratios and direct CP asymmetries. In section \ref{long} we will
propose a model for non-perturbative contribution and compare it 
to experiment. We will then conclude. 

\section{The effective Hamiltonian}
\label{sec:Heff}
The effective weak Hamiltonian for charmless hadronic $B$ decays 
consists of a sum of local operators $Q_i$ multiplied by short-distance 
coefficients $C_i$ given in table \ref{tab:wilco}, and products of elements of the quark mixing matrix, $\lambda_p=V_{pb} V_{ps}^*$ or $\lambda_p'=V_{pb} V_{pd}^*$. Below we will focus on $B\to\ P V$ decays; where $P$ and $V$ hold for
pseudoscalar  and vector mesons respectively. Using the unitarity
relation $-\lambda_t=\lambda_u+\lambda_c$, we write
\begin{equation}\label{Heff}
   {\cal H}_{\rm eff} = \frac{G_F}{\sqrt2} \sum_{p=u,c} \!
   \lambda_p \bigg( C_1\,Q_1^p + C_2\,Q_2^p
   + \!\sum_{i=3,\dots, 10}\! C_i\,Q_i + C_{7\gamma}\,Q_{7\gamma}
   + C_{8g}\,Q_{8g} \bigg) + \mbox{h.c.} \,,
\end{equation}
where $Q_{1,2}^p$ are the left-handed current--current operators arising 
from $W$-boson exchange, $Q_{3,\dots, 6}$ and $Q_{7,\dots, 10}$ are 
QCD and electroweak penguin operators, and $Q_{7\gamma}$ and $Q_{8g}$ 
are the electromagnetic and chromomagnetic dipole operators. They are 
given by

\begin{eqnarray}\label{operateurs}
   Q_1^p &=& (\bar p b)_{V-A} (\bar s p)_{V-A} \,,
    \hspace{2.5cm}
    Q^p_2 = (\bar p_i b_j)_{V-A} (\bar s_j p_i)_{V-A} \,, \nonumber\\
   Q_3 &=& (\bar s b)_{V-A} \sum{}_{\!q}\,(\bar q q)_{V-A} \,,
    \hspace{1.7cm}
    Q_4 = (\bar s_i b_j)_{V-A} \sum{}_{\!q}\,(\bar q_j q_i)_{V-A} \,,
    \nonumber\\
   Q_5 &=& (\bar s b)_{V-A} \sum{}_{\!q}\,(\bar q q)_{V+A} \,, 
    \hspace{1.7cm}
    Q_6 = (\bar s_i b_j)_{V-A} \sum{}_{\!q}\,(\bar q_j q_i)_{V+A} \,,
    \nonumber\\
   Q_7 &=& (\bar s b)_{V-A} \sum{}_{\!q}\,{\textstyle\frac32} e_q 
    (\bar q q)_{V+A} \,, \hspace{1.11cm}
    Q_8 = (\bar s_i b_j)_{V-A} \sum{}_{\!q}\,{\textstyle\frac32} e_q
    (\bar q_j q_i)_{V+A} \,, \nonumber \\
   Q_9 &=& (\bar s b)_{V-A} \sum{}_{\!q}\,{\textstyle\frac32} e_q 
    (\bar q q)_{V-A} \,, \hspace{0.98cm}
    Q_{10} = (\bar s_i b_j)_{V-A} \sum{}_{\!q}\,{\textstyle\frac32} e_q
    (\bar q_j q_i)_{V-A} \,, \nonumber\\
   Q_{7\gamma} &=& \frac{-e}{8\pi^2}\,m_b\, 
    \bar s\sigma_{\mu\nu}(1+\gamma_5) F^{\mu\nu} b \,,
    \hspace{0.81cm}
   Q_{8g} = \frac{-g_s}{8\pi^2}\,m_b\, 
    \bar s\sigma_{\mu\nu}(1+\gamma_5) G^{\mu\nu} b \,,
\end{eqnarray}
where $(\bar q_1 q_2)_{V\pm A}=\bar q_1\gamma_\mu(1\pm\gamma_5)q_2$, 
$i,j$ are colour indices, $e_q$ are the electric charges of the quarks 
in units of $|e|$, and a summation over $q=u,d,s,c,b$ is implied. The
definition of the dipole operators $Q_{7\gamma}$ and $Q_{8g}$ corresponds 
to the sign convention $iD^\mu=i\partial^\mu+g_s A_a^\mu t_a$ for the 
gauge-covariant derivative. The Wilson coefficients are calculated at 
a high scale $\mu\sim M_W$ and evolved down to a characteristic scale 
$\mu\sim m_b$ using next-to-leading order renormalization-group 
equations. The essential problem obstructing the calculation of 
non-leptonic decay amplitudes resides in the evaluation of the hadronic 
matrix elements of the local operators contained in the effective 
Hamiltonian.
\begin{table}
\begin{tabular}{|l|c|c|c|c|c|c|}
NLO & $C_1$ & $C_2$ & $C_3$ & $C_4$ & $C_5$ & $C_6$ \\
\hline
$\mu=m_b/2$ & 1.137 & $-0.295$ & 0.021 & $-0.051$ & 0.010 & $-0.065$ \\
$\mu=m_b$   & 1.081 & $-0.190$ & 0.014 & $-0.036$ & 0.009 & $-0.042$ \\
$\mu=2 m_b$ & 1.045 & $-0.113$ & 0.009 & $-0.025$ & 0.007 & $-0.027$ \\
\hline
 & $C_7/\alpha$ & $C_8/\alpha$ & $C_9/\alpha$ & $C_{10}/\alpha$
 & $C_{7\gamma}^{\rm eff}$ & $C_{8g}^{\rm eff}$ \\
\hline
$\mu=m_b/2$ & $-0.024$           & 0.096 & $-1.325$ & 0.331
 & --- & --- \\
$\mu=m_b$   & $-0.011$           & 0.060 & $-1.254$ & 0.223
 & --- & --- \\
$\mu=2 m_b$ & $\phantom{-}0.011$ & 0.039 & $-1.195$ & 0.144
 & --- & --- \\
\hline\hline
LO & $C_1$ & $C_2$ & $C_3$ & $C_4$ & $C_5$ & $C_6$ \\
\hline
$\mu=m_b/2$ & 1.185 & $-0.387$ & 0.018 & $-0.038$ & 0.010 & $-0.053$ \\
$\mu=m_b$   & 1.117 & $-0.268$ & 0.012 & $-0.027$ & 0.008 & $-0.034$ \\
$\mu=2 m_b$ & 1.074 & $-0.181$ & 0.008 & $-0.019$ & 0.006 & $-0.022$ \\
\hline
 & $C_7/\alpha$ & $C_8/\alpha$ & $C_9/\alpha$ & $C_{10}/\alpha$
 & $C_{7\gamma}^{\rm eff}$ & $C_{8g}^{\rm eff}$ \\
\hline
$\mu=m_b/2$ & $-0.012$           & 0.045 & $-1.358$ & 0.418
 & $-0.364$ & $-0.169$ \\
$\mu=m_b$   & $-0.001$           & 0.029 & $-1.276$ & 0.288
 & $-0.318$ & $-0.151$ \\
$\mu=2 m_b$ & $\phantom{-}0.018$ & 0.019 & $-1.212$ & 0.193
 & $-0.281$ & $-0.136$ \\
\end{tabular}
\centerline{
{\it{\caption{\label{tab:wilco}
Wilson coefficients $C_i$ in the NDR scheme. 
Input parameters are $\Lambda^{(5)}_{\overline{\rm
MS}}=0.225$\,$\mathrm{GeV}$,  
$m_t(m_t)=167$\,$\mathrm{GeV}$, $m_b(m_b)=4.2$\,$\mathrm{GeV}$, $M_W=80.4$\,$\mathrm{GeV}$, $\alpha=1/129$, 
and $\sin^2\!\theta_W=0.23$~\cite{0104110}.}}}}
\end{table}
%
\section{QCD factorization in $\boldsymbol{B \to P V}$ decays}
\label{qcdeff}
\begin{figure}
\epsfxsize=11cm
\centerline{\epsffile{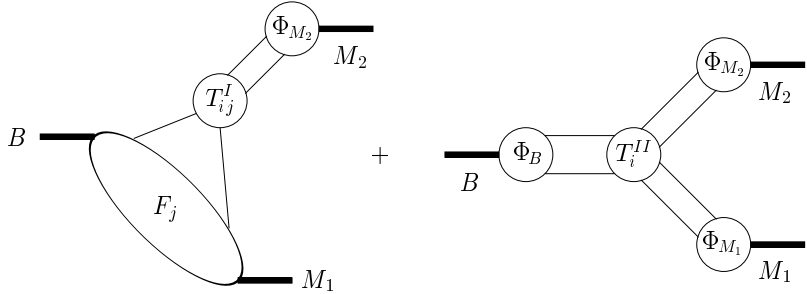}}
\centerline{\parbox{14cm} {\it {\caption{ \label{fig0}
Graphical representation of the
factorization formula. Only one of the two form-factor terms in (\protect\ref{fact}) is shown for simplicity.}}}}
\end{figure}

When the  QCD factorization (QCDF) method is applied to the decays
$B{\to}PV$, the hadronic matrix elements of the local effective operators can be written as

\begin{eqnarray}\label{fact}
   \langle\ P\, V|\Q_i|B\rangle
   &=& F_1^{B\to P}(0)\,T_{V,i}^{\rm I}\star f_V\Phi_V
    + A_0^{B\to V}(0)\,T_{P,i}^{\rm I}\star f_P\Phi_P \nonumber\\
   &&\mbox{}+ T_i^{\rm II} \star f_B\Phi_B \star f_V\Phi_V \star f_P\Phi_P \,,
\end{eqnarray}
where $\Phi_M$ are leading-twist light-cone distribution amplitudes, 
and the $\star$-products imply an integration over the light-cone momentum 
fractions of the constituent quarks inside the mesons. A graphical 
representation of this result is shown in Figure \ref{fig0}.

 Here $F_1^{B{\to}P}$ and $A_0^{B{\to}V}$ denote the form factors for $B{\to}P$
 and $B{\to}V$ transitions, respectively. ${\Phi}_{B}(\xi)$,
 ${\Phi}_{V}(x)$, and ${\Phi}_{P}(y)$ are the light-cone distribution
 amplitudes (LCDA) of valence quark Fock
 states for $B$, vector, and pseudoscalar mesons, respectively.
 $T^{I,II}_{i}$ denote the hard-scattering kernels, which are dominated by
 hard gluon exchange when the power suppressed 
 ${\cal O}({\Lambda}_{QCD}/m_{b})$ terms are neglected. So they are
 calculable order by order in perturbation theory. The leading terms of
 $T^{I}$ come from the tree level and correspond to the naive
 factorization (NF) approximation.
 The order of ${\alpha}_{s}$ terms of $T^I$ can be depicted by 
 vertex-correction diagrams Fig.\ref{fig1} (a-d) and penguin-correction diagrams Fig.\ref{fig1} (e-f). $T^{II}$ describes the hard interactions between the
 spectator quark and the emitted meson $M_{2}$ when the gluon virtuality
 is large. Its lowest order terms are ${\cal O}({\alpha}_{s})$ and can be
 depicted by hard spectator scattering diagrams Fig.\ref{fig1} (g-h). One of the
 most interesting results of the QCDF approach is that, in the heavy quark
 limit, the strong phases arise naturally from the hard-scattering kernels
 at the order of ${\alpha}_{s}$. 
 As for the nonperturbative part, they are, as already mentioned, taken 
 into account by the form factors and the LCDA of mesons up to corrections
 which are power suppressed in $1/m_b$.

 With the above discussions on the effective Hamiltonian of $B$ decays
 Eq.(\ref{Heff}) and the QCDF expressions of hadronic matrix
 elements Eq.(\ref{fact}), the decay amplitudes for $B{\to}P\,\,V$ in the
 heavy quark limit can be written as
 \begin{equation}
 {\cal A}(B{\to}PV) = \frac{G_{F}}{\sqrt{2}}
    \sum\limits_{p=u,c} \sum\limits_{i=1}^{10} \lambda_{p}\,\, a_{i}^{p}\,\,
   {\langle}P\,\, V{\vert}{\cal O}_{i}{\vert}B{\rangle}_{nf}.
 \label{eq:decay-f}
 \end{equation}
 The above ${\langle}PV{\vert}{\cal O}_{i}{\vert}B{\rangle}_{ nf}$ are the 
 factorized hadronic matrix elements, which have the same definitions as
 those in the NF approach. The ``nonfactorizable'' effects are included in
 the coefficients $a_{i}$ which are process dependent. The coefficients
 $a_{i}$ are collected in Sec. \ref{sec23}, and the explicit expressions
 for the decay amplitudes of $B{\to}P\,\,V$ can be found in 
the appendix \ref{sec:app0}.

 According to the arguments in \cite{QCDF}, the contributions of weak
 annihilation to the decay amplitudes are power suppressed, and they
 do not appear in the QCDF formula Eq.(\ref{fact}). But, as emphasized
 in \cite{Nagashima:2002ia,0004173,0004213}, the contributions from weak annihilation
 could give large strong phases with QCD corrections, and hence large CP
 violation could be expected, so their effects cannot simply be neglected.
 However, in the QCDF method, the annihilation topologies (see Fig.\ref{fig2})
 violate factorization because of the endpoint divergence. There is
 similar endpoint divergence when considering the chirally enhanced hard
 spectator scattering. One possible way is to treat the endpoint
 divergence from different sources as different phenomenological
 parameters \cite{0104110}. The corresponding price is the introduction 
 of model dependence and extra numerical uncertainties in the decay
 amplitudes. In this work, we will follow the treatment of Ref.
 \cite{0104110} and express the weak annihilation topological decay
 amplitudes as
 \begin{equation}
 {\cal A}^{a}(B{\to}PV)\,\, {\propto}\,\, f_{B}\,\, f_{P}\,\, f_{V}
    {\sum} \lambda_{p}\,\, b_{i} \; ,
 \label{eq:decay-a}
 \end{equation}
 where the parameters $b_{i}$ are collected in Sec. \ref{sec24}, and the
 expressions for the weak annihilation decay amplitudes of $B{\to}P\,\,V$ are
 listed in the appendix \ref{sec:app1}. 
 \subsection{The QCD coefficients $\boldsymbol{a_{i}}$}
 \label{sec23}
\begin{figure}
 \begin{center}
 \begin{picture}(300,160)(0,30)
 \put(-90,-450){\epsfxsize180mm\epsfbox{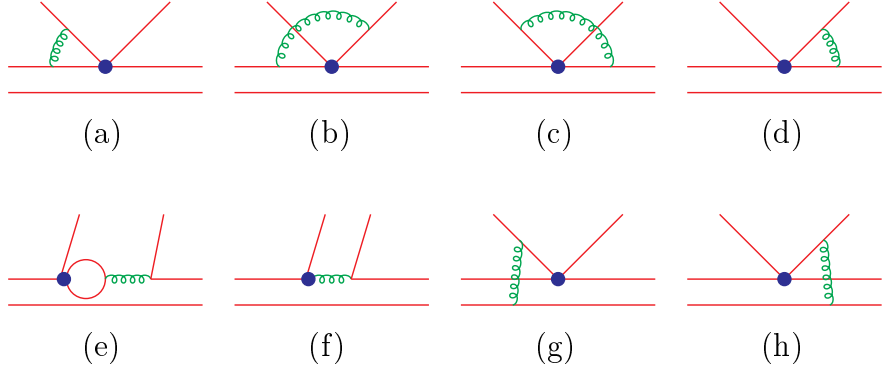}}
 \end{picture}
 \end{center}
 \it{\caption{ \label{fig1}
Order ${\alpha}_{s}$ corrections to the
 hard-scattering kernels. The upward quark lines represent the emitted
 mesons from b quark weak decays. These diagrams are commonly called
 vertex corrections, penguin corrections, and hard spectator
 scattering diagrams for Fig. (a-d), Fig. (e-f), and Fig. (g-h) respectively.}}
 \end{figure}

We express the QCD coefficients  $a_{i}$ (see Eq.(\ref{eq:decay-f}) ) in two parts, i.e.,
$a_{i}=a_{i,I}+a_{i,II}$. The first term $a_{i,I}$ contains the naive
factorisation and  the vertex corrections
 which are described by Fig.\ref{fig1} (a-f), while
the second part $a_{i,II}$ corresponds to the hard spectator
scattering diagrams Fig.\ref{fig1} (g-h).

There are two different cases according to the final states. Case I
is that the recoiled meson $M_{1}$ is a vector meson, and the emitted
meson $M_{2}$ corresponds to a pseudoscalar meson, and vice versa for
case II. For case I, we sum up the results for $a_{i}$ as follows:


 \begin{eqnarray}
&&a_{1,I} = C_{1} + \frac{C_{2}}{N_{c}}
                \Big[
                1 + \frac{C_{F}{\alpha}_{s}}{4{\pi}} V_{M}
                \Big],
      \ \ \ \ \ \ \ \ \ \ \ \ \ \ \ \ \ \ \ \ \ \ \ 
      a_{1,II} = \frac{{\pi}C_{F}{\alpha}_{s}}{N_{c}^{2}}
                 C_{2} H(BM_{1},M_{2}), \nn\\
&&a_{2,I} = C_{2} + \frac{C_{1}}{N_{c}}
                \Big[
                1 + \frac{C_{F}{\alpha}_{s}}{4{\pi}} V_{M}
                \Big],
      \ \ \ \ \ \ \ \ \ \ \ \ \ \ \ \ \ \ \ \ \ \ \ 
      a_{2,II} = \frac{{\pi}C_{F}{\alpha}_{s}}{N_{c}^{2}}
                 C_{1} H(BM_{1},M_{2}),  \nn\\
&&a_{3,I} = C_{3} + \frac{C_{4}}{N_{c}}
                \Big[
                1 + \frac{C_{F}{\alpha}_{s}}{4{\pi}} V_{M}
                \Big],
      \ \ \ \ \ \ \ \ \ \ \ \ \ \ \ \ \ \ \ \ \ \ \ 
      a_{3,II} = \frac{{\pi}C_{F}{\alpha}_{s}}{N_{c}^{2}}
                 C_{4} H(BM_{1},M_{2}),  \nn\\
&&a_{4,I}^{p} = C_{4} + \frac{C_{3}}{N_{c}}
                \Big[
                1 + \frac{C_{F}{\alpha}_{s}}{4{\pi}} V_{M}
                \Big]
                  + \frac{C_{F}{\alpha}_{s}}{4{\pi}}
                    \frac{P_{M,2}^{p}}{N_{c}},
      \ \ \ \ \ 
      a_{4,II} = \frac{{\pi}C_{F}{\alpha}_{s}}{N_{c}^{2}}
                 C_{3} H(BM_{1},M_{2}),  \nn\\
&&a_{5,I} = C_{5} + \frac{C_{6}}{N_{c}}
                \Big[
                1 - \frac{C_{F}{\alpha}_{s}}{4{\pi}} V_{M}^{\prime}
                \Big],
      \ \ \ \ \ \ \ \ \ \ \ \ \ \ \ \ \ \ \ \ \ \ \  
      a_{5,II} = - \frac{{\pi}C_{F}{\alpha}_{s}}{N_{c}^{2}}
                 C_{6} H^{\prime}(BM_{1},M_{2}),  \nn\\
&&a_{6,I}^{p} = C_{6} + \frac{C_{5}}{N_{c}}
                \Big[
                1 - 6 \frac{C_{F}{\alpha}_{s}}{4{\pi}}
                \Big]
                    + \frac{C_{F}{\alpha}_{s}}{4{\pi}}
                      \frac{P_{M,3}^{p}}{N_{c}},
      \ \ \ \ \ \ \ \ 
      a_{6,II} = 0,  \nn\\
&&a_{7,I} = C_{7} + \frac{C_{8}}{N_{c}}
                \Big[
                1 - \frac{C_{F}{\alpha}_{s}}{4{\pi}} V_{M}^{\prime}
                \Big],
      \ \ \ \ \ \ \ \ \ \ \ \ \ \ \ \ \ \ \ \ \ \ \    
      a_{7,II} = - \frac{{\pi}C_{F}{\alpha}_{s}}{N_{c}^{2}}
                 C_{8} H^{\prime}(BM_{1},M_{2}),  \nn\\
&&a_{8,I}^{p} = C_{8} + \frac{C_{7}}{N_{c}}
                \Big[
                1 - 6 \frac{C_{F}{\alpha}_{s}}{4{\pi}}
                \Big]
                    + \frac{\alpha}{9{\pi}}
                      \frac{P_{M,3}^{p,ew}}{N_{c}},
      \ \ \ \ \ \ \ \ \ \ \ 
      a_{8,II} = 0,  \nn\\
&&a_{9,I} = C_{9} + \frac{C_{10}}{N_{c}}
                \Big[
                1 + \frac{C_{F}{\alpha}_{s}}{4{\pi}} V_{M}
                \Big],
      \ \ \ \ \ \ \ \ \ \ \ \ \ \ \ \ \ \ \ \ \ \ \     
      a_{9,II} = \frac{{\pi}C_{F}{\alpha}_{s}}{N_{c}^{2}}
                 C_{10} H(BM_{1},M_{2}),  \nn\\
&&a_{10,I}^{p} = C_{10} + \frac{C_{9}}{N_{c}}
                \Big[
                 1 + \frac{C_{F}{\alpha}_{s}}{4{\pi}} V_{M}
                \Big]
                    + \frac{\alpha}{9{\pi}}
                      \frac{P_{M,2}^{p,ew}}{N_{c}},
      \ \ \ \ \ \  
      a_{10,II} = \frac{{\pi}C_{F}{\alpha}_{s}}{N_{c}^{2}}
                  C_{9} H(BM_{1},M_{2}), \label{eq:ai}
 \end{eqnarray}

\noi where $C_{F}=\frac{N_{c}^{2}-1}{2N_{c}}$, and $N_{c}=3$. The vertex
 parameters $V_{M}$ and $V_{M}^{\prime}$ result from Fig.\ref{fig1}
 (a-d); the QCD penguin parameters $P_{M,i}^{p}$ and the electroweak
 penguin parameters $P_{M,i}^{p,ew}$ result from Fig.\ref{fig1} (e-f). 
 
The vertex corrections are given by:
\begin{eqnarray}\label{FM}
   V_M &=& 12\ln\frac{m_b}{\mu} - 18 + \int_0^1\! dx\,g(x)\,\Phi_M(x)
    \,, \nonumber\\
   V_M' &=& 12\ln\frac{m_b}{\mu} - 6 + \int_0^1\! dx\,g(1-x)\,\Phi_M(x)
    \,, \nonumber\\
   g(x) &=& 3\left( \frac{1-2x}{1-x}\ln x-i\pi \right) \nonumber\\
   &&\mbox{}+ \left[ 2 \,\mbox{Li}_2(x) - \ln^2\!x + \frac{2\ln x}{1-x}
    - (3+2i\pi)\ln x - (x\leftrightarrow 1-x) \right] ,
\end{eqnarray}

\noi where $\mbox{Li}_2(x)$ is the dilogarithm function, whereas the constants $18$ and $6$ are specific to the NDR scheme.
The penguin contributions are:
 \begin{eqnarray}
  P_{M,2}^{p} &=& C_{1} \Big[ \frac{4}{3} {\ln}\frac{m_{b}}{\mu}
                            + \frac{2}{3} - G_{M}(s_{p}) \Big]
                 + \Big( C_{3} - \frac{1}{2} C_{9} \Big)
                   \Big[ \frac{8}{3} {\ln}\frac{m_{b}}{\mu}
                       + \frac{4}{3} - G_{M}(0) - G_{M}(1) \Big]
             \nonumber \\
             & & + \sum\limits_{q=q^{\prime}} \Big( C_{4} + C_{6}
                       + \frac{3}{2} e_{q} C_{8}
                       + \frac{3}{2} e_{q} C_{10} \Big)
                   \Big[ \frac{4}{3} {\ln}\frac{m_{b}}{\mu}
                       - G_{M}(s_{q}) \Big]
             - 2 C_{8g}^{\rm eff} {\int}_{0}^{1}dx \
               \frac{{\Phi}_{M}(x)}{1-x}, \nn\\
  P_{M,3}^{p} &=& C_{1} \Big[ \frac{4}{3} {\ln}\frac{m_{b}}{\mu}
                            + \frac{2}{3} - {\hat{G}}_{M}(s_{p}) \Big]
                 + \Big( C_{3} - \frac{1}{2} C_{9} \Big)
                   \Big[ \frac{8}{3} {\ln}\frac{m_{b}}{\mu}
                       + \frac{4}{3} - {\hat{G}}_{M}(0)
                       - {\hat{G}}_{M}(1) \Big]
             \nonumber \\
             & & + \sum\limits_{q=q^{\prime}} \Big( C_{4} + C_{6}
                       + \frac{3}{2} e_{q} C_{8}
                       + \frac{3}{2} e_{q} C_{10} \Big)
                   \Big[ \frac{4}{3} {\ln}\frac{m_{b}}{\mu}
                       - {\hat{G}}_{M}(s_{q}) \Big]
                       - 2 C_{8g}^{\rm eff},\label{eq:penguin} 
\end{eqnarray}
\ and the electroweak penguin parameters $P_{M,i}^{p,ew}$:

\begin{eqnarray}
  P_{M,2}^{p,ew} &=& \Big( C_{1} + N_{c} C_{2} \Big)
                     \Big[ \frac{4}{3} {\ln}\frac{m_{b}}{\mu}
                         + \frac{2}{3} - G_{M}(s_{p}) \Big]
                - \Big( C_{3} + N_{c} C_{4} \Big)
                     \Big[ \frac{4}{3} {\ln}\frac{m_{b}}{\mu}
                         + \frac{2}{3} - \frac{1}{2} G_{M}(0)
                         - \frac{1}{2} G_{M}(1) \Big]
               \nonumber \\
               & & + \sum\limits_{q=q^{\prime}} \Big( N_{c} C_{3} + C_{4}
                   + N_{c} C_{5} + C_{6} \Big) \frac{3}{2} e_{q}
                     \Big[ \frac{4}{3} {\ln}\frac{m_{b}}{\mu}
                         - G_{M}(s_{q}) \Big]
                - N_{c} C_{7{\gamma}}^{\rm eff} {\int}_{0}^{1}dx \
                     \frac{{\Phi}_{M}(x)}{1-x}, \nn\\
  P_{M,3}^{p,ew} &=& \Big( C_{1} + N_{c} C_{2} \Big)
                     \Big[ \frac{4}{3} {\ln}\frac{m_{b}}{\mu}
                         + \frac{2}{3} - {\hat{G}}_{M}(s_{p}) \Big]
                - \Big( C_{3} + N_{c} C_{4} \Big)
                     \Big[ \frac{4}{3} {\ln}\frac{m_{b}}{\mu}
                         + \frac{2}{3} - \frac{1}{2} {\hat{G}}_{M}(0)
                          - \frac{1}{2} {\hat{G}}_{M}(1) \Big]
               \nonumber \\
               & & + \sum\limits_{q=q^{\prime}} \Big( N_{c} C_{3} + C_{4}
                   + N_{c} C_{5} + C_{6} \Big) \frac{3}{2} e_{q}
                     \Big[ \frac{4}{3} {\ln}\frac{m_{b}}{\mu}
                         - {\hat{G}}_{M}(s_{q}) \Big]
                   - N_{c} C_{7{\gamma}}^{\rm eff}, \label{eq:ewpenguin}
 \end{eqnarray}
 
\noi where $s_{q}=m_{q}^{2}/m_{b}^{2}$, and where $q^{\prime}$ in 
the expressions for
 $P_{M,i}^{p}$ and $P_{M,i}^{p,ew}$  runs over all
 the active quarks at the  scale ${\mu}={\cal O}(m_{b})$, i.e.,
 $q^{\prime}=u,d,s,c,b$. The functions $G_M(s)$ and  $\hat G_M(s)$ are
 given respectively by:

\begin{eqnarray}
G_M(s) &=& \int_0^1\!dx\,G(s-i\epsilon,1-x)\,\Phi_M(x) \,, \label{GK}\\
\hat G_M(s) &=& \int_0^1\!dx\,G(s-i\epsilon,1-x)\,\Phi_M^p(x) \,,
\label{penfunction1}\\  
G(s,x) &=& -4\int_0^1\!du\,u(1-u) \ln[s-u(1-u)x] \nonumber\\
&=& \frac{2(12s+5x-3x\ln s)}{9x}
- \frac{4\sqrt{4s-x}\,(2s+x)}{3x^{3/2}}
\arctan\sqrt{\frac{x}{4s-x}} \,. 
\end{eqnarray}

The parameters $H(BM_{1},M_{2})$ and $H^{\prime}(BM_{1},M_{2})$ in 
 $a_{i,II}$, which originate from hard gluon exchanges between the
 spectator quark and the emitted meson $M_{2}$, are written as:
\begin{eqnarray}
  H(BV,P)&=&\frac{f_{B}f_{V}}{m_{B}^{2}A_{0}^{B{\to}V}(0)}
  {\int}_{0}^{1} d{\xi} {\int}_{0}^{1} dx {\int}_{0}^{1} dy
   \frac{{\Phi}_{B}({\xi})}{\xi}
   \frac{{\Phi}_{P}(x)}{\bar{x}}
   \frac{{\Phi}_{V}(y)}{\bar{y}},
   \nn\\
  H^{\prime}(BV,P)&=&\frac{f_{B}f_{V}}{m_{B}^{2}A_{0}^{B{\to}V}(0)}
  {\int}_{0}^{1} d{\xi} {\int}_{0}^{1} dx {\int}_{0}^{1} dy
   \frac{{\Phi}_{B}({\xi})}{\xi}
   \frac{{\Phi}_{P}(x)}{x}
   \frac{{\Phi}_{V}(y)}{\bar{y}}. \label{eq:hard-spectator-p}
 \end{eqnarray}

 For case II (vector meson emitted) except for the parameters of $H(BM_{1},M_{2})$ and
 $H^{\prime}(BM_{1},M_{2})$, the expressions for $a_{i}$ are similar to
 those in case I. However we would like to point out that, because
 ${\langle}V{\vert}(\bar{q}q)_{S{\pm}P}{\vert}0{\rangle}=0$, the
 contributions of the effective operators ${\cal O}_{6,8}$ to the hadronic
 matrix elements vanish, i.e., the terms that are related to $a_{6,8}$
 disappear from the decay amplitudes for case II. As to the parameters
 $H(BM_{1},M_{2})$ and $H^{\prime}(BM_{1},M_{2})$ in $a_{i,II}$, they
 are defined as

 \begin{eqnarray}
  H(BP,V)&=&\frac{f_{B}f_{P}}{m_{B}^{2}F_{1}^{B{\to}P}(0)}
  {\int}_{0}^{1} d{\xi} {\int}_{0}^{1} dx {\int}_{0}^{1} dy
   \frac{{\Phi}_{B}({\xi})}{\xi}
   \frac{{\Phi}_{V}(x)}{\bar{x}} 
   \Big[ \frac{{\Phi}_{P}(y)}{\bar{y}}
       + \frac{2 {\mu}_{P}}{m_b} \frac{\bar{x}}{x}
         \frac{{\Phi}_{P}^{p}(y)}{\bar{y}} \Big],
   \nn\\
  H^{\prime}(BP,V)&=& - \frac{f_{B}f_{P}}{m_{B}^{2}F_{1}^{B{\to}P}(0)}
  {\int}_{0}^{1} d{\xi} {\int}_{0}^{1} dx {\int}_{0}^{1} dy
   \frac{{\Phi}_{B}({\xi})}{\xi}
   \frac{{\Phi}_{V}(x)}{x} 
   \Big[ \frac{{\Phi}_{P}(y)}{\bar{y}}
       + \frac{2 {\mu}_{P}}{m_b} \frac{x}{\bar{x}}
         \frac{{\Phi}_{P}^{p}(y)}{\bar{y}} \Big].
 \label{eq:hard-spectator-v}
\end{eqnarray}
\noi The parameter $\mu_{P}=m_P^2/(m_1+m_2)$ where $m_{1,2}$ are 
the current quark masses of the meson constituents, is proportional 
the the chiral quark condensate.

 \subsection{The annihilation parameters $\boldsymbol{b_{i}}$}
 \label{sec24}
 The parameters of $b_{i}$ in Eq.(\ref{eq:decay-a}) correspond to weak
 annihilation contributions. Now we give their expressions, which are
 analogous to those in \cite{0104110}:

 \begin{eqnarray}
  & &b_{1}(M_{1},M_{2}) = \frac{C_{F}}{N_{c}^{2}} C_{1}
                           A_{1}^{i}(M_{1},M_{2}),\nn\\
  & &b_{2}(M_{1},M_{2}) = \frac{C_{F}}{N_{c}^{2}} C_{2}
                           A_{1}^{i}(M_{1},M_{2}), \nn\\
  & &b_{3}(M_{1},M_{2}) = \frac{C_{F}}{N_{c}^{2}} \Big\{ C_{3}
                           A_{1}^{i}(M_{1},M_{2})
                   + C_{5} A_{3}^{i}(M_{1},M_{2})
             + \Big[ C_{5} + N_{c} C_{6} \Big]
                           A_{3}^{f}(M_{1},M_{2}) \Big\}, \nn\\
  & &b_{4}(M_{1},M_{2}) = \frac{C_{F}}{N_{c}^{2}} \Big\{ C_{4}
                           A_{1}^{i}(M_{1},M_{2})
                   + C_{6} A_{2}^{i}(M_{1},M_{2}) \Big\}, \nn\\
  & &b_{3}^{ew}(M_{1},M_{2}) = \frac{C_{F}}{N_{c}^{2}} \Big\{ C_{9}
                           A_{1}^{i}(M_{1},M_{2})
                   + C_{7} A_{3}^{i}(M_{1},M_{2})
             + \Big[ C_{7} + N_{c} C_{8} \Big]
                           A_{3}^{f}(M_{1},M_{2}) \Big\}, \nn\\
  & &b_{4}^{ew}(M_{1},M_{2}) = \frac{C_{F}}{N_{c}^{2}} \Big\{ C_{10}
                           A_{1}^{i}(M_{1},M_{2})
                   + C_{8} A_{2}^{i}(M_{1},M_{2}) \Big\}. \label{eq:bi}
 \end{eqnarray} 

\begin{figure}[t]
 \begin{center}
 \begin{picture}(300,100)(0,30) 
 \put(-90,-500){\epsfxsize180mm\epsfbox{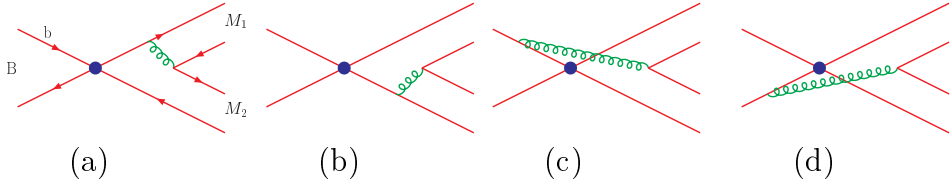}} 
 \end{picture}
 \end{center}
\it{ \caption{\label{fig2}
Order ${\alpha}_{s}$ corrections to the weak
annihilations of charmless decays $B{\to}P\,V$.}}
\end{figure}

 Here the current-current annihilation parameters $b_{1,2}(M_{1},M_{2})$
 arise from the hadronic matrix elements of the effective operators
 ${\cal O}_{1,2}$, the QCD penguin annihilation parameters $b_{3,4}(M_{1},M_{2})$ from ${\cal O}_{3-6}$, and the electroweak penguin annihilation parameters $b_{3,4}^{ew}(M_{1},M_{2})$ from ${\cal O}_{7-10}$. The parameters of $b_{i}$
 are closely related to the final states; they can also be divided into
 two different cases according to the final states. Case I is that $M_{1}$
 is a vector meson, and $M_{2}$ is a pseudoscalar meson (here $M_{1}$ and
 $M_{2}$ are tagged in Fig.~\ref{fig2}). Case II is that $M_{1}$ corresponds to a
 pseudoscalar meson, and $M_{2}$ corresponds to a vector meson. For
 case I, the definitions of $A^{i,f}_{k}(M_{1},M_{2})$
 in Eq.(\ref{eq:bi}) are

 \begin{eqnarray}
 & &A^{f}_{1,2}(V,P) = 0,\nn\\
 & &A^{f}_{3}(V,P) = {\pi}{\alpha}_{s} {\int}_{0}^{1} dx {\int}_{0}^{1} dy
            {\Phi}_{V}(x) {\Phi}_{P}^{p}(y) \frac{2 {\mu}_{P}}{m_b}
             \frac{2 (1+\bar{x})}{{\bar{x}}^{2}y},\nn\\
 & &A^{i}_{1}(V,P) = {\pi}{\alpha}_{s}
            {\int}_{0}^{1} dx {\int}_{0}^{1} dy
            {\Phi}_{V}(x) {\Phi}_{P}(y)
       \Big[ \frac{1}{y(1-x\bar{y})}
           + \frac{1}{{\bar{x}}^{2}y} \Big],\nn\\
 & &A^{i}_{2}(V,P) = - {\pi}{\alpha}_{s}
            {\int}_{0}^{1} dx {\int}_{0}^{1} dy
            {\Phi}_{V}(x) {\Phi}_{P}(y)
       \Big[ \frac{1}{\bar{x}(1-x\bar{y})}
           + \frac{1}{\bar{x}y^{2}} \Big],\nn\\
 & &A^{i}_{3}(V,P) = {\pi}{\alpha}_{s}
            {\int}_{0}^{1} dx {\int}_{0}^{1} dy
            {\Phi}_{V}(x) {\Phi}_{P}^{p}(y)
             \frac{2 {\mu}_{P}}{m_b}
             \frac{2\bar{y}}{\bar{x}y(1-x\bar{y})}.
 \label{eq:bi-ivp-3}
 \end{eqnarray}

 For case-II,
 \begin{eqnarray}
 & &A^{f}_{1,2}(P,V)=0,\nn\\
 & &A^{f}_{3}(P,V)= - {\pi}{\alpha}_{s} {\int}_{0}^{1} dx
            {\int}_{0}^{1} dy {\Phi}_{P}^{p}(x) {\Phi}_{V}(y)
             \frac{2 {\mu}_{P}}{m_b} \frac{2 (1+y)}{\bar{x}y^{2}},\nn\\
 & &A^{i}_{1}(P,V) = {\pi}{\alpha}_{s}
            {\int}_{0}^{1} dx {\int}_{0}^{1} dy
            {\Phi}_{P}(x) {\Phi}_{V}(y)
       \Big[ \frac{1}{y(1-x\bar{y})}
           + \frac{1}{{\bar{x}}^{2}y} \Big],\nn\\
 & &A^{i}_{2}(P,V) = - {\pi}{\alpha}_{s}
            {\int}_{0}^{1} dx {\int}_{0}^{1} dy
            {\Phi}_{P}(x) {\Phi}_{V}(y)
       \Big[ \frac{1}{\bar{x}(1-x\bar{y})}
           + \frac{1}{\bar{x}y^{2}} \Big],\nn\\
 & &A^{i}_{3}(P,V) = {\pi}{\alpha}_{s}
            {\int}_{0}^{1} dx {\int}_{0}^{1} dy
            {\Phi}_{P}^{p}(x) {\Phi}_{V}(y)
             \frac{2 {\mu}_{P}}{m_b}
             \frac{2x}{\bar{x}y(1-x\bar{y})}.
 \label{eq:bi-ipv}
 \end{eqnarray}

 Here our notation and convention are the same as those in \cite{0104110}.
 The superscripts $i$ and $f$ on $A^{i,f}$ correspond to the contributions
 from Fig.~\ref{fig2}(a-b) and Fig.\ref{fig2}(c-d), respectively. The subscripts $k=1,2,3$ on
 $A^{i,f}_{k}$ refer to the Dirac structures $(V-A){\otimes}(V-A)$,
 $(V-A){\otimes}(V+A)$ and $(-2)(S-P){\otimes}(S+P)$, respectively.
 ${\Phi}_{V}(x)$ denotes the leading-twist LCDAs of a vector meson, and
 ${\Phi}_{P}(x)$ and ${\Phi}_{P}^{p}(x)$ denote twist-2 and twist-3 LCDAs
 of a pseudoscalar meson, respectively.

 Note that assuming $SU(3)$ flavor symmetry implies symmetric 
 LCDAs of light mesons (under
 $x{\leftrightarrow}\bar{x}$), whence
 $A_{1}^{i}=-A_{2}^{i}$. In this approximation the weak annihilation
 contributions (for case I) can be parametrized as
\begin{eqnarray}\label{eq:aif-kvp}
 A_{1}^{i}(V,P) &{\simeq}& 18{\pi}{\alpha}_{s}\Big(X_{A}-4
             +\frac{{\pi}^{2}}{3}\Big), \nn\\
 A_{3}^{i}(V,P) &{\simeq}& {\pi}{\alpha}_{s}r_{\chi}\bigg[
              2{\pi}^{2}-6\Big(X_{A}^{2}+2X_{A}\Big)\bigg], \nn\\
 A_{3}^{f}(V,P) &{\simeq}& 6{\pi}{\alpha}_{s}r_{\chi}
             \Big(2X_{A}^{2}-X_{A}\Big),
 \end{eqnarray}
where $X_{A}={\int}^{1}_{0}dx/x$ parametrizes the divergent endpoint
 integrals and $r_{\chi}=2 \mu_{P}/m_b$ is the so-called {\it chirally enhanced} factor. We can get similar forms to Eq.(\ref{eq:aif-kvp}) for case II,
 but with $A_{3}^{f}(P,V)=-A_{3}^{f}(V,P)$. In our calculation, we will
 treat $X_{A}$ as a phenomenological parameter, and take the same value
 for all annihilation terms, although this approximation is crude and
 there is no known physical argument justifying this assumption. We
 shall see below that $X_{A}$ gives large uncertainties in the theoretical
 prediction.

\section{ QCD factorisation  versus experiment}
\label{exp}
In order to propose a test of QCD factorization with respect to
experiment, a compilation of various charmless branching fractions and
direct $CP$ asymmetries was performed and is given in tables
\ref{tab:BRexp}, \ref{tab:CPexp} and \ref{tab:correl}. This
compilation includes the latest results from BaBar, Belle and CLEO.
The measurements were combined into a single central value and error,
that may be compared with the theoretical prediction. First, the total
error from each experiment was computed by summing quadratically the
statistic and systematic error: this approach is valid in the limit
that the systematic error is not so large with respect to the
statistic error. Secondly, when the experiment provides an asymmetric
error ${}^{+\sigma_1}_{-\sigma_2}$, a conservative symmetric error was
assumed in the calculation by using $\pm
\mathrm{Max}(\sigma_1,\sigma_2)$. In case of a disagreement between
several experiments for a given
measurement, the total error was increased by a ``scale
factor'' computed from a $\chi^2$ combining the various experiments,
using the standard procedure given by the PDG \cite{PDG}.

In order to compare the theoretical predictions $\{y\}$ with the
experimental measurements $\{x\pm \sigma_x\}$, the following $\chi^2$
was defined:
$$
\chi^2=\sum\left(\frac{x-y}{\sigma_x}\right)^2.
$$
In the case when a correlation matrix between several measurements
is given by the experiment, as in the case of the $\rho^+\pi^-/\rho^+
K^-$ measurements, the $\chi^2$ was corrected to account for it.  The
above $\chi^2$ was then minimized using MINUIT~\cite{MINUIT}, letting
free all theoretical parameters in their allowed range. 
The quality of the minimum yielded by MINUIT was assessed by replacing it
with an ad hoc minimizer scanning the entire parameter space.
The theoretical predictions,  with the theoretical parameters yielding 
the best fits, are compared to
experiment in table \ref{tab:fit} for two scenarios to be explained below. 
 The asymmetries of the $\rho^\pm\pi^\mp$  channels can be 
 expressed~\cite{Aubert:2002jq} 
 in terms of the quantities reported in table \ref{tab:correl}. The comparison 
 between their theoretical predictions and experiment is reported in  
  table~\ref{tab:asym}.
 
Scenario 1 refers to a fit according to QCD factorisation,
varying all theoretical parameters in the range presented in table
\ref{tab:input}.  Even the unitarity triangle angle $\gamma$ is varied
freely and ends up not far from 90 degrees.  We have taken $X_A=X_H$
in the range proposed in ref.~\cite{0104110}:
\begin{equation}\label{eq:xa}
X_{A,H}=\int \limits_0^1 \frac{dx}{x} = \ln \frac{m_B}{\Lambda_{h}}
(1+\rho_{A,H}\,e^{i\phi_{A,H}}).
\end{equation}
These parameters label our ignorance of the non perturbatively
calculable subdominant contribution to the annihilation and hard
scattering, defined in Eqs.~(\ref{eq:bi-ivp-3}, \ref{eq:bi-ipv})
and Eqs.~(\ref{eq:hard-spectator-p}, \ref{eq:hard-spectator-v}) respectively.
  They do not need to have the same value for all $PV$
channels but we have nevertheless assumed one common value since a fit
would become impossible with too many unknown parameters.

Scenario 2 in table \ref{tab:fit}  refers to  a fit adding  a charming penguin
inspired long distance contribution which will be presented and discussed in  
section~\ref{long}. In this fit $\gamma$ is constrained within the range
$[34^\circ, 82^\circ]$. 

The values of the theoretical parameters found for the two best fits
is given in table \ref{tab:input}: many parameters are found to be at
the edge of their allowed range\footnote{Table 5 shows that the fit 
value of $\rho_A$ appears at the edge of the input range, $\rho_A=1$. 
However enlarging the range of $\rho_A$, such  as $|\rho_A|\leq 10$, 
brings a large annihilation contributions  $(\rho_A,\phi_A )= 
(2.3, -41^{o})$ for scenario 1 and $(4.4, -108^{o})$ for scenario 2. With so large
values of $|\rho_A|$ the unpredictable contributions would dominate the
total result making the whole exercize void of signification.}. 
In order to estimate the quality of
the agreement between measurements and predictions, the standard
Monte Carlo based ``goodness of fit'' test was performed:
\begin{itemize}
\item the best-fit values of the theoretical parameters were used to
  make predictions for the branching ratios and $CP$ asymmetries,
\item the total experimental error from each measurement was used to
  generate new experimental values distributed around the predictions
  with a Gaussian probability,
\item the full fit previously performed on real measurements is now
  run on this simulated data, and the $\chi^2$ of this fit is saved in
  a histogram $H$.
\end{itemize}
It is then possible to compare the $\chi^2_{\mathrm{data}}$ obtained
from the measurement with the $\chi^2$ one would obtain if the
predictions were true. Additionally, one may compute the confidence
level of the tested model by using:
$$
CL \le \frac{\int_{\chi^2>\chi^2_{\mathrm{data}}} H(\chi^2) d\chi^2}
{\int_{\chi^2>0}  H(\chi^2) d\chi^2}.
$$

\begin{figure}
  \psfrag{chi2}[tr][tr]{\large $\chi^2$}
  \psfrag{experiments}[br][br]{\large experiments}
  \begin{center}
  \subfigure[Scenario 1]{\includegraphics[width=7cm]{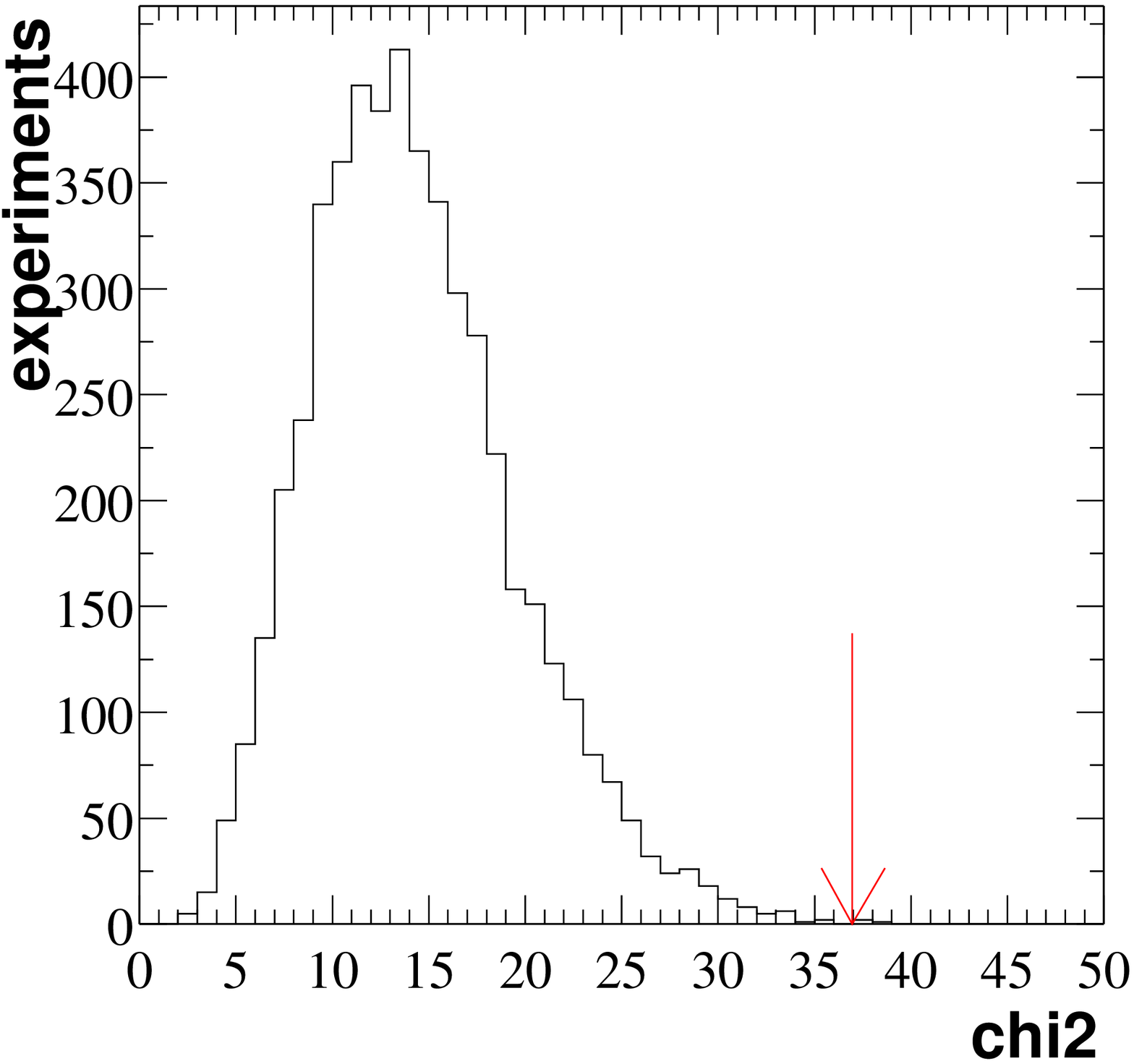}}
  \hspace{5mm}
  \subfigure[Scenario 2]{\includegraphics[width=7cm]{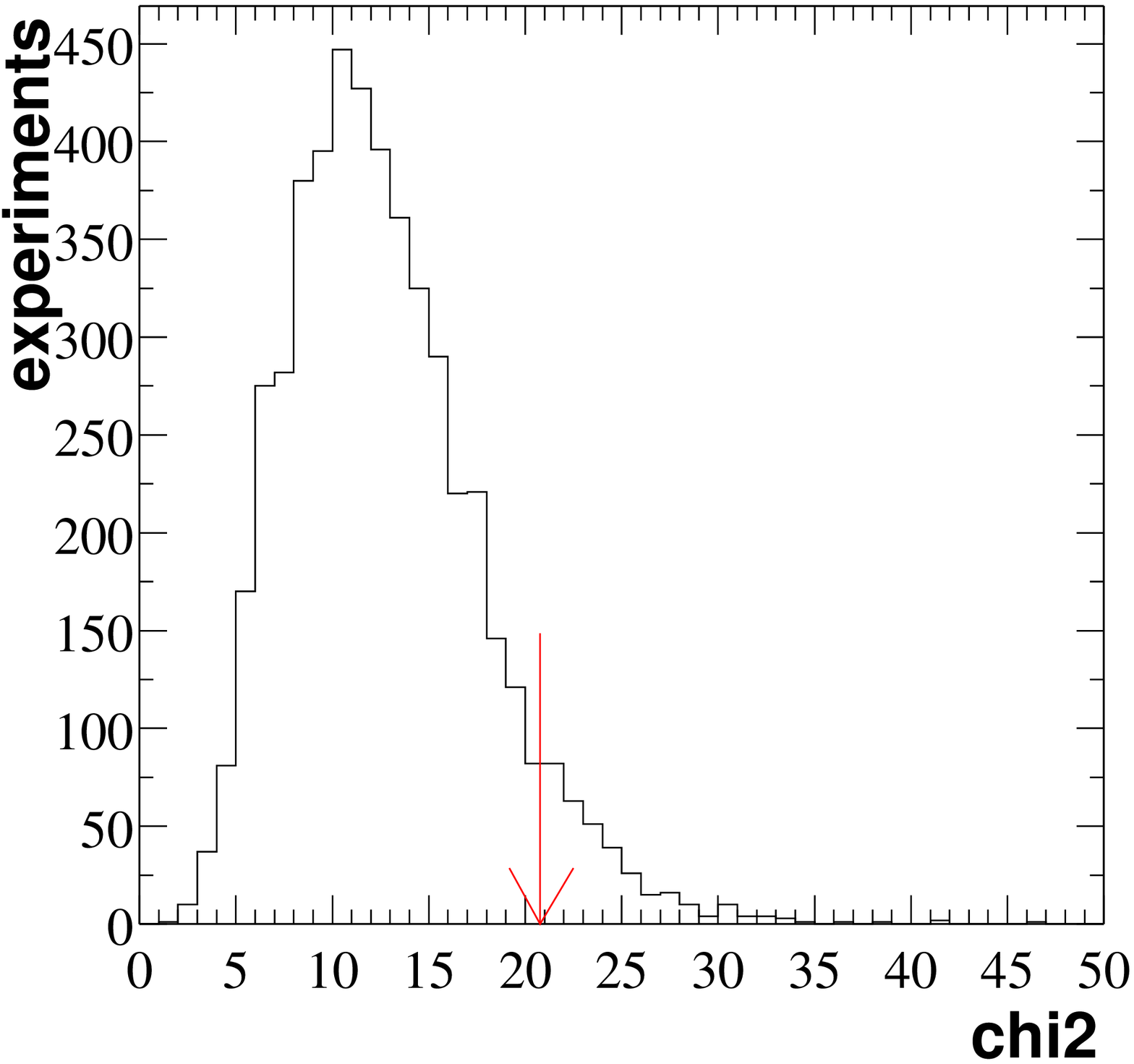}}
  \end{center}
  \it{\caption{Goodness of fit test of the two proposed theoretical
    models: the arrow points at the value $\chi^2_{\mathrm{data}}$
    found from the measurements, and the histogram shows the values
    allowed for $\chi^2$ in the case that the models predictions are
    correct.}}
  \label{fig:goodness}
\end{figure}

The results of the ``goodness of fit'' tests are given in
\figurename~\ref{fig:goodness}. From these tests, one may quote an
upper limit for the confidence level in scenario 1, $CL \le 0.1\%$, and
in the case of scenario 2, $CL \le 7.7\%$.

\renewcommand{\arraystretch}{1.3}
\begin{table}[htb]
\begin{center}
\begin{tabular}{|c|c|c|c|c|}
${\mathcal{BR}}$($\times 10^6$)&  BaBar\citer{0107058,BaBar} &
Belle\citer{Belle,0201007} &
CLEO\citer{0101032,CLEO} & Average  \\ \hline
$B^0 \to \pi^\pm \rho^\mp$
& $28.9 \pm 5.4 \pm 4.3$
& $20.8 ^{+6.0}_{-6.3}{}^{+2.8}_{-3.1}$
& $27.6^{+8.4}_{-7.4} \pm 4.2$
& $25.53 \pm 4.32$
\\ \hline
$B^+ \to \pi^+ \rho^0$
& $24 \pm 8 \pm 3 (<39)$
& $8.0^{+2.3}_{-2.0} \pm 0.7$
& $10.4^{+3.3}_{-3.4} \pm 2.1$
& $9.49 \pm 2.57$
\\ \hline
$B^0 \to \pi^0 \rho^0$
& $3.6\pm 3.5\pm 1.7(<10.6)$
& $<5.3$
& $1.6^{+2.0}_{-1.4}\pm 0.8(<5.5)$
& $2.07\pm 1.88$
\\ \hline
$B^+ \to \pi^+ \omega$
& $6.6^{+2.1}_{-1.8}\pm 0.7$
& $4.2^{+2.0}_{-1.8}\pm 0.5$
& $11.3^{+3.3}_{-2.9}\pm 1.4$
& $6.22\pm 1.70$
\\ \hline
$B^0 \to K^+ \rho^-$
& --
& $15.8^{+5.1}_{-4.6}{}^{+1.7}_{-3.0}$
& $16.0^{+7.6}_{-6.4}\pm 2.8(<32)$
& $15.88\pm 4.65$
\\ \hline
$B^+ \to K^+ \rho^0$
& $10\pm 6\pm 2(<29)$
& --
& $8.4^{+4.0}_{-3.4}\pm 1.8(<17)$
& $8.92\pm 3.60$
\\ \hline
$B^+ \to K^+ \omega$
& $1.4^{+1.3}_{-1.0}\pm 0.3(<4)$
& $9.2^{+2.6}_{-2.3}\pm 1.0$
& $3.2^{+2.4}_{-1.9}\pm 0.8(<7.9)$
& $2.92\pm 1.94$
\\ \hline
$B^0 \to  K^0 \omega$
& $5.9^{+1.7}_{-1.5}\pm 0.9$
& --
& $10.0^{+5.4}_{-4.2}\pm 1.4 (<21)$
& $6.34\pm 1.82$
\\ \hline
$B^0 \to  K^{\ast\,+} \pi^-$
& --
& $26.0\pm 8.3\pm 3.5$
& $16^{+6}_{-5}\pm 2$
& $19.3\pm 5.2$
\\ \hline
$B^+ \to  K^{\ast\,0} \pi^-$
& $15.5\pm 3.4\pm 1.8$
& $19.4^{+4.2}_{-3.9}\pm 2.1^{+3.5}_{-6.8}$
& $7.6^{+3.5}_{-3.0}\pm 1.6(<16)$
& $12.12\pm 3.13$
\\ \hline
$B^+ \to  K^{\ast\,-} \pi^0$
& --
& --
& $7.1^{+11.4}_{-7.1}\pm 1.0(<31)$
& $7.1\pm 11.4$
\\ \hline
$B^+ \to  K^{\ast\,+} \eta$
& $22.1^{+11.1}_{-9.2}\pm 3.3$
& $26.5^{+7.8}_{-7.0}\pm 3.0$
& $26.4^{+9.6}_{-8.2}\pm 3.3$
& $25.4\pm 5.6$
\\ \hline
$B^0 \to  K^{\ast\,0} \eta$
& $19.8^{+6.5}_{-5.6}\pm 1.7$
& $16.5^{+4.6}_{-4.2}\pm 1.2$
& $13.8^{+5.5}_{-4.6}\pm 1.6$
& $16.41\pm 3.21$
\\ \hline
$B^+ \to  K^+  \phi$
& $9.2\pm 1.0 \pm 0.8$
& $10.7\pm 1.0^{+0.9}_{-1.6}$
& $5.5^{+2.1}_{-1.8}\pm 0.6$
& $8.58\pm 1.24$
\\ \hline
$B^0 \to  K^0  \phi$
& $8.7^{+1.7}_{-1.5}\pm 0.9$
& $10.0^{+1.9}_{-1.7}{}^{+0.9}_{-1.3}$
& $5.4^{+3.7}_{-2.7}\pm 0.7(<12.3)$
& $8.72\pm 1.37$
\\ 
\end{tabular}
\end{center}
\centerline{
{\it{
\caption{\label{tab:BRexp} Experimentally known data of CP-averaged 
branching ratios for
the charmless $B\to PV$ decay modes, used as input for the global fit.
The channels containing the $\eta'$ meson have been excluded.}}}}
\end{table}


\renewcommand{\arraystretch}{1.3}
\begin{table}[htb]
\begin{center}
\begin{tabular}{|c|c|c|c|c|}
${\mathcal A_{CP}}$&  BaBar\citer{0107058,BaBar} &
Belle\citer{Belle,0201007} &
CLEO\citer{0101032,CLEO} & Average  \\ \hline
$B^- \to \pi^- \omega$
& $-0.01^{+0.29}_{-0.31}\pm 0.03$
& --
& $-0.34\pm 0.25\pm 0.02$
& $-0.21\pm 0.19$
\\ \hline
$B^- \to K^- \omega$
& --
& $-0.21\pm 0.28\pm 0.03$
& --
& $\-0.21\pm 0.28$
\\ \hline
$B^- \to  K^{\ast\,-} \eta$
& --
& $-0.05^{+0.25}_{-0.30}\pm 0.01$
& --
& $-0.05\pm 0.30$
\\ \hline
${\overline B}^0 \to {\overline K}^{\ast\,0} \eta$
& --
& $0.17^{+0.28}_{-0.25}\pm 0.01$
& --
& $0.17\pm 0.28$
\\ \hline
$B^- \to  K^-  \phi$
& $-0.05\pm 0.20\pm 0.03$
& --
& --
& $-0.05\pm 0.20$
\\ 
\end{tabular}
\end{center}
\centerline{
\parbox{14cm}
{\it{
\caption{\label{tab:CPexp}Experimental measured data of direct 
CP asymmetries for
the charmless $B\to PV$ decay modes, used as input for the global fit.
}}}}
\end{table}

\begin{table}[htb!]
\begin{center}
\begin{tabular}{|c|c|cccc|}
\multirow{2}{1.9cm}{$B^0 \to \rho^\pm \pi^\mp$}
& \multirow{2}{2cm}{Measurement}
& \multicolumn{4}{c|}{Correlation coefficient (\%)}
\\ \cline{3-6}
&& ${\mathcal A^{\rho\pi}_{CP}}$
& ${\mathcal A^{\rho K}_{CP}}$
& ${\mathcal C_{\rho\pi}}$
& $\Delta {\mathcal C_{\rho\pi}}$
\\ \hline
${\mathcal A^{\rho\pi}_{CP}}$
& $-0.22\pm 0.08\pm 0.07$
& --
& $3.4$
& $-11.8$
& $-10.4$
\\
${\mathcal A^{\rho K}_{CP}}$
& $0.19\pm 0.14\pm 0.11$
& $3.4$
& --
& $-1.3$
& $-1.1$
\\
${\mathcal C_{\rho\pi}}$
& $0.45^{+0.18}_{-0.19}\pm 0.09$
& $-11.8$
& $-1.3$
& --
& $23.9$
\\
$\Delta {\mathcal C_{\rho\pi}}$
& $0.38^{+0.19}_{-0.20}\pm 0.11$
& $-10.4$
& $-1.1$
& $23.9$
& --
\\
\end{tabular}
\end{center}
\centerline{\parbox{10cm}{\it{
\caption{\label{tab:correl}Experimental results and correlation matrix 
for the various asymmetries
measured in the channels $\rho^\pm \pi^\mp/\rho^\pm K^\mp$. 
The notations are explained in  \cite{Aubert:2002jq}.}}}}
\end{table}

In tables \ref{tab:BRexp}  (\ref{tab:CPexp}) we give the experimental 
CP-averaged branching ratios (direct CP asymmetries) which we have used in
 our fits. We have also used the quantities reported in table \ref{tab:correl}
which are related to the  branching ratios and CP asymmetries of the
$B\to \rho^\pm\pi^\mp$ channels.
 
\renewcommand{\arraystretch}{1.3}
\begin{table}[htb!]
\begin{center}
  \begin{tabular}{|l|c|c|c|}
Input & Range & Scenario 1 & Scenario 2 \\ \hline
$\gamma$ (deg)&  & $99.955$ & $81.933$\\
$m_s$ (GeV)& $[0.085,0.135]$ & $0.085$ & $0.085$\\
$\mu$ (GeV)& $[2.1,8.4]$ & $3.355$ & $5.971$\\
$\rho_A$& $[-1,1]$ & $1.000$ & $1.000$\\
$\phi_A$(deg)& $[-180,180]$ & $-22.928$ & $-87.907$\\
$\lambda_B$ (GeV)& $[0.2,0.5]$ & $0.500$ & $0.500$\\
$f_B$   (GeV)& $[0.14,0.22]$ & $0.220$ & $0.203$\\
$R_u$& $[0.35,0.49]$ & $0.350$ & $0.350$\\
$R_c$& $[0.018,0.025]$ & $0.018$ & $0.018$\\
$A_0^{B\to \rho}$& $[0.3162,0.4278]$ & $0.373$ & $0.377$\\
$F_1^{B\to \pi}$& $[0.23,0.33]$ & $0.330$ & $0.301$\\
$A_0^{B\to \omega}$& $[0.25,0.35]$ & $0.350$ & $0.326$\\
$A_0^{B\to K^*}$& $[0.3995,0.5405]$ & $0.400$ & $0.469$\\
$F_1^{B\to K}$& $[0.28,0.4]$ & $0.333$ & $0.280$\\
${\mathrm Re}[{\mathcal A}^{P}]$& $[-0.01,0.01]$ & & $0.00253$\\
${\mathrm Im}[{\mathcal A}^{P}]$& $[-0.01,0.01]$ & & $-0.00181$\\
${\mathrm Re}[{\mathcal A}^{V}]$& $[-0.01,0.01]$ & & $-0.00187$\\
${\mathrm Im}[{\mathcal A}^{V}]$& $[-0.01,0.01]$ & & $0.00049$\\
  \end{tabular}
\end{center}
\centerline{\parbox{10cm}{\it{
\caption{\label{tab:input}Various theoretical inputs used in our global
analysis of $B\,\to\,PV$  decays in QCDF. The parameter
ranges have been taken from
literature~\cite{0104110},\cite{Yang:2000xn},\cite{Du:2002up} and
~\cite{Ball:1998kk}. The two last columns give the best fits of both scenarios.}}}}
\end{table}

\renewcommand{\arraystretch}{1.3}
\begin{table}
\begin{center}
  \begin{tabular}{|l|c|cc|cc|}
 & Experiment & \multicolumn{2}{c|}{Scenario 1} & \multicolumn{2}{c|}{Scenario 2} \\ 
 & & Prediction & $\chi^2$ & Prediction & $\chi^2$ \\ \hline
${\mathcal BR}({\overline B}^0\,\to\,\rho^0\,\pi^0)$ & $2.07 \pm 1.88$ & 0.132 & 1.1 & 0.177 & 1.0\\
${\mathcal BR}({\overline B}^0\,\to\,\rho^+\,\pi^-)$ &  & 11.023 &  & 10.962 & \\
${\mathcal BR}({\overline B}^0\,\to\,\rho^-\,\pi^+)$ &  & 18.374 &  & 17.429 & \\
${\mathcal BR}({\overline B}^0\,\to\,\rho^{\pm}\,\pi^{\mp})$ & $25.53 \pm 4.32$ & 29.397 & 0.8 & 28.391 & 0.4\\
${\mathcal BR}(B^-\,\to\,\rho^0\,\pi^-)$ & $9.49 \pm 2.57$ & 9.889 & 0.0 & 7.879 & 0.4\\
${\mathcal BR}(B^-\,\to\,\omega\,\pi^-)$ & $6.22 \pm 1.7$ & 6.002 & 0.0 & 5.186 & 0.4\\
${\mathcal BR}(B^-\,\to\,K^{*-}\,K^0)$ &  & 0.457 &  & 0.788 & \\
${\mathcal BR}(B^-\,\to\,K^{*0}\,K^-)$ &  & 0.490 &  & 0.494 & \\
${\mathcal BR}(B^-\,\to\,\Phi\,\pi^-)$ &  & 0.004 &  & 0.003 & \\
${\mathcal BR}(B^-\,\to\,\rho^-\,\pi^0)$ &  & 9.646 &  & 11.404 & \\
${\mathcal BR}({\overline B}^0\,\to\,\rho^0\,{\overline K}^0)$ &  & 5.865 &  & 8.893 & \\
${\mathcal BR}({\overline B}^0\,\to\,\omega\,{\overline K}^0)$ & $6.34 \pm 1.82$ & 2.318 & 4.9 & 5.606 & 0.2\\
${\mathcal BR}({\overline B}^0\,\to\,\rho^+\,K^-)$ & $15.88 \pm 4.65$ & 6.531 & 4.0 & 14.304 & 0.1\\
${\mathcal BR}({\overline B}^0\,\to\,K^{*-}\,\pi^+)$ & $19.3 \pm 5.2$ & 9.760 & 3.4 & 10.787 & 2.7\\
${\mathcal BR}( B^-\,\to\,K^{*-}\,\pi^0)$ & $7.1 \pm 11.4$ & 7.303 & 0.0 & 8.292 & 0.0\\
${\mathcal BR}({\overline B}^0\,\to\,\Phi\,{\overline K}^0)$ & $8.72 \pm 1.37$ & 8.360 & 0.1 & 8.898 & 0.0\\
${\mathcal BR}(B^-\,\to\,{\overline K}^{*0}\,\pi^-)$ & $12.12 \pm 3.13$ & 7.889 & 1.8 & 11.080 & 0.1\\
${\mathcal BR}(B^-\,\to\,\rho^0\,K^-)$ & $8.92 \pm 3.6$ & 1.882 & 3.8 & 5.655 & 0.8\\
${\mathcal BR}(B^-\,\to\,\rho^{-}\,{\overline K}^0)$ &  & 7.140 &  & 14.006 & \\
${\mathcal BR}(B^-\,\to\,\omega\,K^-)$ & $2.92 \pm 1.94$ & 2.398 & 0.1 & 6.320 & 3.1\\
${\mathcal BR}(B^-\,\to\,\Phi\,K^-)$ & $8.88 \pm 1.24$ & 8.941 & 0.0 & 9.479 & 0.2\\
${\mathcal BR}({\overline B}^0\,\to\,{\overline K}^{*0}\,\eta)$ & $16.41 \pm 3.21$ & 22.807 & 4.0 & 18.968 & 0.6\\
${\mathcal BR}(B^-\,\to\,K^{*-}\,\eta)$ & $25.4 \pm 5.6$ & 17.855 & 1.8 & 15.543 & 3.1\\
$\Delta\,{\mathcal C_{\rho\pi}}$ & $0.38 \pm 0.23$ & 0.250 & \multirow{4}{1.2cm}{$\left. \begin{matrix} \\ \\ \\ \\
\end{matrix}\right\}8.1/4$} & 0.228 & \multirow{4}{1.2cm}{$\left. \begin{matrix} \\ \\ \\ \\
\end{matrix}\right\}3.9/4$}\\
${\mathcal C_{\rho\pi}}$ & $0.45 \pm 0.21$ & 0.019 &  & 0.092 & \\
${\mathcal A_{CP}^{\rho\,\pi}}$ & $-0.22 \pm 0.11$ & -0.015 &  & -0.115 & \\
${\mathcal A_{CP}^{\rho\,K}}$ & $0.19 \pm 0.18$ & 0.060 &  & 0.197 & \\
${\mathcal A_{CP}^{\omega\,\pi^-}}$ & $-0.21 \pm 0.19$ & -0.072 & 0.5 & -0.198 & 0.0\\
${\mathcal A_{CP}^{\omega\,K^-}}$ & $-0.21 \pm 0.28$ & 0.029 & 0.7 & 0.189 & 2.0\\
${\mathcal A_{CP}^{\eta\,K^{*-}}}$ & $-0.05 \pm 0.3$ & -0.138 & 0.1 & -0.217 & 0.3\\
${\mathcal A_{CP}^{\eta\,{\overline K}^{*0}}}$ & $0.17 \pm 0.28$ & -0.186 & 1.6 & -0.158 & 1.4\\
${\mathcal A_{CP}^{\phi\,K^{-}}}$ & $-0.05 \pm 0.2$ & 0.006 & 0.1 & 0.005 & 0.1\\
\cline{4-4}\cline{6-6}
 &  & & 36.9 & & 20.8\\  
  \end{tabular}
\centerline{\parbox{10cm}{\it{
\caption{\label{tab:fit}
Best fit values using the global analysis of $B\,\to\,PV$
decays in QCDF with free $\gamma$ (scenario 1) and QCDF+Charming Penguins 
(scenario 2) with constrained $\gamma$. The CP-averaged branching ratios are in unit of $10^{-6}$.}}}}
\vspace*{1.cm}
\begin{tabular}{|c|c|c|c|}
  & Experiment & Scenario 1 & Scenario 2\\ \hline
${\mathcal A_{CP}^{\rho^+\,\pi^-}}$ & $-0.82\pm 0.31\pm 0.16$ & -0.04 & -0.23 \\
\hline
${\mathcal A_{CP}^{\rho^-\,\pi^+}}$ & $-0.11\pm 0.16\pm 0.09$ & -0.0002 & 0.04 \\
\end{tabular}
\end{center}
\centerline{\parbox{10cm}{\it{
\caption{\label{tab:asym}Values of the CP asymmetries for $B\,\to\,\pi\rho$
decays in QCDF (scenario 1) and QCDF+Charming Penguins (scenario 2).
The notations are explained in  \cite{Aubert:2002jq}.}}}}
\end{table}
  
For the sake of definiteness let us remind that the branching ratios for any
 charmless $B$ decays, $B\to P V$ channel,
 in the rest frame of the $B$-meson, is given by
\beq\label{eq:Br}
{\mathcal BR}(B\to P V)=\frac{\tau_B}{8 \pi}\frac{|p|}{m_B^2}
|{\cal A}(B\to P V) + {\cal A}^{a}(B\to P V) + {\cal A}^{\rm LD}(B\to P V)|^2,
\eeq

\noi where $\tau_B$  represents the
$B$-meson life time (charged or uncharged according to the case).
The amplitudes ${\cal A}, {\cal A}^{a}$ and ${\cal A}^{\rm LD}$
are defined in appendix A, B and in eqs.~(\ref{eq:LDd}) and (\ref{eq:LDs}) respectively. 
In the case of pure QCD factorisation (scenario 1) we take of course
${\cal A}^{\rm LD} = 0$.  The
kinematical factor $|p|$ is written as:
\beq\label{eq:p}
|p|=\frac{\sqrt{[m_B^2-(m_P+m_V)^2][m_B^2-(m_P-m_V)^2]}}{2 m_B}.
\eeq

\subsection{Comparison  with Du \it{et\ al}}
Our negative conclusion about the QCD factorisation  fit
of the $B\to PV$ channels is at odds with the conclusion of the authors
of ref. \cite{Du:2002cf}, who have performed a successful fit of 
both $B\to PP$ and $B\to PV$ channels using the same theoretical starting point.
These authors have excluded from their fits the channels containing a 
$K^\ast$ in the final state, arguing that these channels seemed questionable to
them. We have thus made a fit without the channels containing the $K^\ast$,
 and indeed we find as the authors of ref.~\cite{Du:2002cf}
that the global agreement between QCD factorisation and experiment was
satisfactory. Notice that this fit was done without discarding  
the channels $B^+\to \omega \pi^+(K^+)$ as done by Du {\it et al}.

Notice also that the parameters $C_{\rho\pi}$ and 
the  $A^{\rho\pi}_{CP}$ have been kept in this fit.
The disagreement between QCDF and experiment for these quantities was
not enough to spoil the satisfactory agreement of the global fit
 because the experimental errors are still large on these quantities.  

The conclusion of this subsection is that the  difference between the
``optimistic'' conclusion about QCDF  of Du {\it et al} and our rather
pessimistic one comes from their choice of discarding  the channels containing
the $K^\ast$'s. In other words the conclusion about the status of QCDF in the
$B\to PV$ channels depends on the confidence  we give to the published results on
these channels.
   
 \section{A simple model for long distance interactions}
 \label{long}
As seen in table \ref{tab:fit} the failure of our overall fit with 
QCDF can be traced to two main facts.
First the strange branching ratios are underestimated by QCDF. Second the 
direct CP asymmetries in the non-strange channels might also be underestimated. 
A priori this could be cured if some non-perturbative mechanism 
was contributing to $|P|$. Indeed, first, in the strange channels, $|P|$
 is Cabibbo 
enhanced and such a non-perturbative contribution could increase 
the branching ratios,
and second, increasing $|P|/|T|$ in the non-strange channels with non-small
strong phases could increase significantly the  direct CP asymmetries
as already discussed. We have therefore tried a charming-penguin inspired  
 model. We wanted nevertheless to avoid to add too many new parameters
 which would make the fit void of signification. We have therefore 
 tried a model for long distance penguin contributions which depends only
 on two fitted complex numbers. 
  
 Let us start by describing our charming-penguin inspired  
 model for strange final states.
  In the ``charming penguin'' picture the weak decay of
 a $\overline B^0$ ($B^-$) meson through the action of the 
 operator $Q_1^c$ (see notations in Eqs.~(\ref{Heff})
 and~(\ref{operateurs})) creates an hadronic system containing 
 the quarks $s, \bar d (\bar u), c, \bar c$, for example 
 $\overline D_s^{(\ast)}$ + $D^{(\ast)}$ systems. This system goes to
 long distances, the $ c, \bar c$ eventually annihilate, a pair
 of light quarks are created by non-perturbative strong interaction 
 and one is left with  two light meson. Let us here restrict ourselves 
  to the case of a $P V$ pair of mesons, i.e. one of the 
 final mesons is a light pseudoscalar ($\pi, K, \eta$)
  and the other a light vector meson ($\rho, \omega, \phi, K^\ast$).
    We leave aside from now on the $\eta'$ which is presumably quite
  special. 
  
 We will picture  now this hadronic system as a coherent state which decays 
 into the two final mesons with total strangeness -1. This state  
 has a total angular momentum $J=0$. Its flavor
  $s \bar d$ is that of a member of an octet of flavor-$SU(3)$. We will
   assume  flavor-$SU(3)$ symmetry in the decay amplitude of this 
   hadronic state. This still leaves four $SU(3)$-invariant amplitudes
   since both $P$ and $V$ can have an octet and a singlet component
   and that there exist two octets in the decomposition of  $8\times 8$.
   We make a further simplifying assumption based on the OZI rule.
   Let us give an example:
   we assume that $V= (s \bar q)$ where $q$ is any of the light quarks $u,d,s$,
   and that $P=(q \bar d)$. Then we compute the contractions between
   \beq\label{LD1}
   <(s \bar q)(q \bar d) | s (\bar u u + \bar d d +\bar s s) \bar d > = 1
   \eeq

The meaning of this rule is simple. We add to the $s \bar d$ quarks 
in our hadronic state an $SU(3)$ singlet $\bar u u + \bar d d +\bar s s$
and compute an ``overlap'' making contractions so that the quarks in the singlet go into two
different mesons. This latter constraint is the OZI rule. 
This is why the overlap in Eq.~(\ref{LD1}) is 1 even if $q=d$ since it is
forbidden to have both $d$ quarks  from the singlet in the same final meson.
As an example, the decay $B \to \overline K^{0} \rho^0$
gives the following overlap coefficient:
   \beq\label{overlap}
   <(s \bar d)\frac{(u \bar u - d \bar d)}{\sqrt 2} | 
   s (\bar u u + \bar d d +\bar s s) \bar d >  = - \frac  1{\sqrt 2}
   \eeq
   
For the $\eta$ meson we will use the decomposition in \cite{ali}.
   The overlap coefficients thus computed play the role of $SU(3)$ 
   Clebsch-Gordan (CG) coefficients  computed in a simple way. These
    coefficients are assumed to
   be multiplied by an universal complex amplitude to be fitted from experiment.
  Up to now we have assumed that the active quark (here $s$) ended up in the
  vector meson.  We need another universal amplitude for the case where  the
  active quark ends up in the pseudoscalar meson.

   We are thus left with two theoretically independent and unknown
   amplitudes, one with  $V= (s \bar q),\, P=(q \bar d)$,
  one with $P= (s \bar q),\, V=(q \bar d)$. 
   We  shall write them
  respectively as ${\cal A^{P}}$  (${\cal A^{V}}$) when 
  the active quark ends up in the Pseudoscalar (Vector) meson.

  Concerning the $\overline B$ decay into a pseudoscalar + vector meson of
  vanishing total  strangeness, we apply the same recipe with the 
  same amplitudes ${\cal A^{P}}$   and ${\cal A^{V}}$, replacing the $s$ quark
   by a $d$ quark and, of course, the corresponding replacement of 
  the CKM factor $V_{cb}V_{cs}^\ast$ by $V_{cb}V_{cd}^\ast$.

 \paragraph{ To summarize} : the long distance term is given by two universal
   complex amplitudes multiplied by a CG coefficient computed simply by the overlap factor
   in (\ref{overlap}), see table \ref{tab:clebsch}.
\vspace*{1.cm}
\begin{table}[b]
\begin{center}
\begin{tabular}{|l|c|c|}
  & $Cl^P$ &  $Cl^V$  \\ 
 \hline
${\mathcal BR}({\overline B}^0\,\to\,\rho^0\,\pi^0)$ & 0.5 & 0.5 \\
${\mathcal BR}({\overline B}^0\,\to\,\rho^+\,\pi^-)$ & 1.0 & 0. \\
${\mathcal BR}({\overline B}^0\,\to\,\rho^-\,\pi^+)$ & 0. & 1.0  \\
${\mathcal BR}(B^-\,\to\,\rho^0\,\pi^-)$ & $1/\sqrt 2$ & $-1/\sqrt 2$\\
${\mathcal BR}(B^-\,\to\,\omega\,\pi^-)$ & $1/\sqrt 2$ & $1/\sqrt 2$\\
${\mathcal BR}(B^-\,\to\,K^{*-}\,K^0)$ &  1.0 & 0. \\
${\mathcal BR}(B^-\,\to\,K^{*0}\,K^-)$ &  0. & 1.0  \\
${\mathcal BR}(B^-\,\to\,\Phi\,\pi^-)$ & 0. & 0. \\
${\mathcal BR}(B^-\,\to\,\rho^-\,\pi^0)$ & $-1/\sqrt 2$ & $1/\sqrt 2$\\
${\mathcal BR}({\overline B}^0\,\to\,\rho^0\,{\overline K}^0)$ & $-1/\sqrt 2$ & 0.\\
${\mathcal BR}({\overline B}^0\,\to\,\omega\,{\overline K}^0)$  & $1/\sqrt 2$ & 0.\\
${\mathcal BR}({\overline B}^0\,\to\,\rho^+\,K^-)$ & 1.0 & 0. \\
${\mathcal BR}({\overline B}^0\,\to\,K^{*-}\,\pi^+)$& 0. & 1.0  \\
${\mathcal BR}( B^-\,\to\,K^{*-}\,\pi^0)$ & 0. & $1/\sqrt 2$ \\
${\mathcal BR}({\overline B}^0\,\to\,\Phi\,{\overline K}^0)$ & 0. & 1.0  \\
${\mathcal BR}(B^-\,\to\,{\overline K}^{*0}\,\pi^-)$ & 0. & 1.0  \\
${\mathcal BR}(B^-\,\to\,\rho^0\,K^-)$ & $1/\sqrt 2$ & 0.\\
${\mathcal BR}(B^-\,\to\,\rho^{-}\,{\overline K}^0)$ & 1.0 & 0. \\
${\mathcal BR}(B^-\,\to\,\omega\,K^-)$ & $1/\sqrt 2$ & 0.\\
${\mathcal BR}(B^-\,\to\,\Phi\,K^-)$  & 0. & 1.0  \\
${\mathcal BR}({\overline B}^0\,\to\,{\overline K}^{*0}\,\eta)$ &-0.665 &0.469 \\
${\mathcal BR}(B^-\,\to\,K^{*-}\,\eta)$ &-0.665 &0.469 
\end{tabular}
\end{center}
\centerline{
{\it{
\caption{\label{tab:clebsch} Flavor-$SU(3)$  Clebsch-Gordan coefficient
for long distance penguin-like contributions. Notice that the channel
$B^-\,\to\,\Phi\,\pi^-$ vanishes due to OZI rule.  }}}}
\end{table}

 In practice, to the amplitudes described in the appendices we add the long
 distance  amplitudes, given by:
 \beq\label{eq:LDd}
 {\cal A}^{\rm LD}(B\to PV) = \frac{G_F}{\sqrt{2}}m_B^2 \lambda^\prime_c
 (Cl^P\,{\cal A}^P  +Cl^V\, {\cal A}^V)
 \eeq
 for the non strange channels and 
 \beq\label{eq:LDs}
 {\cal A}^{\rm LD}(B\to PV) = \frac{G_F}{\sqrt{2}}m_B^2 \lambda_c
 (Cl^P\,{\cal A}^P  +Cl^V\, {\cal A}^V)
 \eeq
 for the  strange channels.
 In equations (\ref{eq:LDd}) and (\ref{eq:LDs}), ${\cal A}^P$ and ${\cal A}^V$
 are two complex numbers which are fitted in the global fit of scenario~2 and
 $Cl^P$ and  $Cl^V$ are the flavor-$SU(3)$ Clebsch-Gordan coefficients which are 
 given in table \ref{tab:clebsch}. For both channels containing the $\eta$ 
 we have used the formulae
 \beq
 {Cl}^V =  \frac{{\mathrm cos}\theta_8} {\sqrt{6}}-\frac{{\mathrm
sin}\theta_0}{\sqrt{3}}\qquad {Cl}^P = -2 \,\frac{{\mathrm cos}\theta_8}
 {\sqrt{6}} - \frac{{\mathrm sin}\theta_0}{\sqrt{3}}
 \eeq
 with $\theta_0 = -9.1^\circ$ and $\theta_8 = -22.2^\circ$.

 The fit with long distance penguin contributions is presented in table 
 \ref{tab:fit} under the label ``Scenario 2''. The agreement with experiment is
 improved, it should be so, but not in such a fully convincing manner. The
 goodness of the fit is about 8 \% which implies that this model is not excluded by
 experiment.  However
 a look at table \ref{tab:input} shows that several fitted parameters are still
 stuck at the end of the allowed range of variation. In particular
 $\rho_A = 1$  means that the uncalculable subleading contribution to QCDF 
 is again stretched to its extreme.
 
  Finally the fitted complex numbers
  which fix the size of the long distance penguin contribution (last four lines
  in table \ref{tab:input}) are small. To make this statement quantitative,
  assuming the long distance amplitude were alone, the values for 
  ${\mathcal A}^{P}$ and ${\mathcal A}^{V}$
in table~\ref{tab:fit} correspond to branching ratios which reach at their maximum
 $6 \times 10^{-6}$ but are more generally in the vicinity of  $2\times 10^{-6}$. 
  In part, this is due to the fact that,
  if some strange channels want a large non-perturbative contribution to 
  increase their branching ratios, some other strange channels and 
  particularly the $B \to K \phi$ channels which are in good agreement with QCDF 
  cannot accept the addition of a too large non-perturbative penguin contribution. 
  This last point should be stressed: if the strange channels show a general
  tendency to be underestimated by QCDF, there is the striking exception
  of the $\bar s s s$ 
  channels which agree very well with QCDF and make the case for charming 
  penguins rather difficult.


\section{Conclusion}
\label{conclusion}
We have made a global fit according to QCD factorisation 
of published experimental data concerning charmless 
$B\to PV$ decays including CP asymmetries. We have only excluded 
 from the fit the channels containing the $\eta'$ meson.
 Our conclusion is that it is impossible to reach a good fit. 
As can be seen in the scenario 1 of table \ref{tab:fit}, the reasons
of this failure is that 
the branching ratios for the strange channels are predicted 
significantly smaller than experiment except for the $B\to \phi K$
channels, and in table \ref{tab:asym} 
it can be seen that the direct CP asymmetry of $\overline B\to \rho^+\pi^-$
is predicted very small while experiment gives it very large but only two sigmas 
from zero. Not only is the ``goodness of the fit''
smaller than .1 \%, but the fitted parameters show a tendency to  evade
the allowed domain of QCD factorisation.  One might wonder if we were
not too strict in imposing the same scale $\mu$ in all terms since 
the value of $\mu$, representing the effect of unknown higher
order corrections, could be different in different classes of 
channels~\footnote{We thank Gerhard Buchalla for raising this question.}.
We have performed several tests relaxing this unicity of $\mu$ and 
concluded that it affected very little the outcome of our fit.

For the sake of comparison with the authors of ref.~\cite{Du:2002cf}
we have tried a fit without the channels containing a $K^\ast$. 
The result improves significantly. The only lesson we can receive from
this is that one must look carefully at the evolution of the experimental 
results, many of them being recent, before drawing a final conclusion. 

Both the small predicted branching ratios of the strange channels
and the small predicted direct CP asymmetries in the non strange channels
could be blamed on too small $P$ amplitudes
with too small ``strong phases'' relatively to the $T$ amplitudes. 
We have therefore tried the addition of two ``charming penguin'' inspired 
long distance complex amplitudes combined, in order to make the model
predictive enough, with exact flavor-$SU(3)$ and OZI rule. 
This fit is better than the pure QCDF one: with a 
goodness of the fit of about 8 \% the model is not excluded by experiment.
But the parameters show again a 
tendency to reach the limits of the allowed domain and the best fit 
gives rather small value to the long distance contribution. The latter fact
 is presumably due to the $B\to \phi K$ which are well predicted by QCDF 
 and thus deliver a message which contradicts the other strange channels.
This seems to be the reason of the moderate success of 
our ``charming penguin'' inspired model. 

Altogether, the present situation is unpleasant. QCDF  seems to be unable
 to comply to experiment. QCDF implemented by an 
ad-hoc long distance model is not fully convincing. No clear
hint for the origin of this problem is provided by the total set of
experimental  data.  PQCD, also called  $k_T$ factorisation, 
would predict larger direct CP asymmetries, but we do not know
if their sign would fit experiment neither if an overall agreement
of the branching ratios with data can be achieved.

Maybe  however, the coming experimental data will move enough
to resolve, at least partly, this discrepancy.
We would like to insist on the crucial importance of direct CP asymmetries
in non-strange channels. If they confirm the tendency to be large, 
this would make the case for QCDF really difficult. 

Finally we do not know yet the answer to our initial question:
 are we in a good position to study the unitarity-triangle angle $\alpha$ 
 from indirect CP asymmetries thanks to small penguins. If
 experimental data evolve so as to provide a better support to QCDF, 
 one could become bold enough to use it in estimating $\alpha$ and this would
 reduce the errors. Else, only model-independant bounds\cite{jerome} could be used
 but they are not very constraining in part because of discrete ambiguities.  

 \section*{Acknowledgements}
We are grateful to Gerhard Buchalla for useful discussions on various
aspects of this work. A.~S.~S. would like to thank PROCOPE 2002 and DESY for
 financial support.  We thank Andr\'e Gaidot, Alain Le~Yaouanc, Lluis Oliver, 
 Jean-Claude Raynal and Christophe Y\`eche for several enlightening 
 discussions which initiated this work as well as S\'ebastien Descotes 
 at a later stage. 
\section*{Appendix}
\begin{appendix}
\section{The decay amplitudes for $B{\to}PV$}
\label{sec:app0}\hspace*{\parindent}
Following ref.\cite{ali}, we give the decay amplitudes 
for the following  $B{\to}PV$ decay processes:\\
{\underline{\bf (1) $b\to d$  processes:}}

\begin{eqnarray}
{\cal A} ({\overline{B}}^{0} \to \rho^-  \pi^+)  &=& \frac{G_F}{\sqrt{2}}
m_B^2 f_\rho F_1^{B\to \pi } (m_\rho^2)\Big \{ 
\lambda_u' a_1 +(\lambda_u' +  \lambda_c' )[a_4+a_{10}]\Big \}.\\
{\cal A} ({\overline{B}}^{0} \to \rho^+  \pi^-)  &=& {G_F \over \sqrt{2}} m_B^2 
f_\pi A_0^{B \to \rho }(m_\pi^2) \Big \{ \lambda_u'a_1 +(\lambda_u' +  \lambda_c' )[a_4+a_{10}- r^{\pi}_{\chi}(a_6+a_8)] \Big \}.\\
{\cal A} ({\overline{B}}^{0} \to \pi^0 \rho^0 )&=&-\frac{G_F}{2
\sqrt{2}} m_B^2 \Big(
f_\pi A_0^{B\to \rho }(m_\pi^2)\left \{ \lambda_u'\,\, a_2 
-(\lambda_u' +  \lambda_c' )\left[a_4-\frac{1}{2}a_{10}-
r^{\pi}_{\chi}(a_6-\frac{1}{2}a_8)+ \frac{3}{2}(a_7-a_9)\right]
\right \}  \nonumber \\
&& \Bigl. +f_\rho F_1^{B\to\pi } (m_\rho^2)
\left \{ \lambda_u'a_2 -(\lambda_u' +  \lambda_c' )[a_4-\frac{1}{2}a_{10}-\frac{3}{2}(a_7+a_9)]\right \} \Bigr).\\ 
{\cal A} ( B^- \to \pi^- \rho^0 )&= &\frac{G_F}{2 } m_B^2 \left (
f_\pi A_0^{B\to \rho }(m_\pi^2)
\left \{ \lambda_u'a_1 
+(\lambda_u' +  \lambda_c')[a_4+a_{10}-r^{\pi}_{\chi} (a_6+a_8)] \right \}\right . \nn\\
 &&+ \left . f_\rho F_1^{B\to \pi } (m_\rho^2)
\left \{ \lambda_u' a_2
+(\lambda_u' +  \lambda_c')[-a_4+\frac{1}{2}a_{10}+\frac{3}{2}(a_7+a_9)]\right \}\right).\\ 
{\cal A} ( B^- \to \rho^- \pi^0 )&=&\frac{G_F}{2 } m_B^2\Big(
f_\pi A_0^{B\to \rho }(m_\pi^2) \left \{ \lambda_u'~ a_2 
+(\lambda_u' +  \lambda_c')\left[-a_4+\frac{1}{2}a_{10}
-r^{\pi}_{\chi}(-a_6+\frac{1}{2}a_8)+\frac{3}{2}(a_9-a_7)\right]\right\} \nonumber\\
&& +  \Bigl. f_\rho F_1^{B\to \pi } (m_\rho^2)
\left \{ \lambda_u' ~ a_1 
+(\lambda_u' +  \lambda_c') ~ [a_4+a_{10}]\right \} \Bigr).\\
{\cal A} ( B^- \to \pi^- \omega)  &= &\frac{G_F}{2 } m_B^2\Bigl(
f_\pi A_0^{B\to \omega }(m_\pi^2)   \left \{ \lambda_u'a_1
+(\lambda_u' +  \lambda_c') [a_4+a_{10}-r^{\pi}_{\chi}(a_6+a_8)]\right \}\Bigr. \nn\\
&&+   \left . f_\omega F_1^{B\to \pi } (m_\omega^2)
\left \{ \lambda_u' a_2 +(\lambda_u' +  \lambda_c')\left[a_4+2(a_3+a_5)+\frac{1}{2}(a_7+a_9-a_{10})\right] \right \} \right) .
\end{eqnarray}
 \newpage \pagestyle{plain}
{\underline{\bf (2) $b\to s$  processes:}}
\begin{eqnarray}
  {\cal A} ({\overline{B}}^{0} \to K^{*-} \pi^+)
  &=&\frac{G_F}{\sqrt{2}} m_B^2 f_{K^*} F_1^{B\to \pi }(m_{K^*}^2)
  \Big \{ \lambda_u a_1+(\lambda_u +  \lambda_c) [a_4+a_{10}] \Big \} .\\
{\cal A} ({\overline{B}}^{0} \to K^{-} \rho^+ ) &=& \frac{G_F}{\sqrt{2}} m_B^2 
f_K A_0^{B\to \rho }(m_K^2)\Big \{ \lambda_u a_1 +(\lambda_u +  \lambda_c) [a_4+a_{10}-r^{K}_{\chi}(a_6+a_8)]\Big \}.\\
{\cal A} ({\overline{B}}^{0} \to {\overline K}^0 \rho^0)  & =&  \frac{G_F}{2} m_B^2
\left \{f_{K} A_0^{B\to \rho } (m_{K^{0}}^2) (-\lambda_u -  \lambda_c)
\left[a_4-\frac{1}{2}a_{10} 
-r^{K}_{\chi}(a_6-\frac{1}{2}a_8)\right]\right.\nonumber\\
 &&+ \left.  f_\rho F_1^{B\to K }(m_\rho^2)\left [ \lambda_u a_2 
+(\lambda_u +\lambda_c)\times\frac{3}{2}(a_9+a_{7}) \right ]\right\}.\\
{\cal A} ( B^- \to K^{*- }\pi^0)  &= & \frac{G_F}{2} m_B^2\left [
f_\pi A_0^{B\to K^* }(m_\pi^2)
\left \{ \lambda_u a_2 +(\lambda_u +\lambda_c)\times\frac{3}{2}
(a_9-a_{7})\right \} \right .\nonumber\\
&&+  \left . f_{K^*} F_1^{B\to \pi } (m_{K^{*}}^2) \left \{
\lambda_u a_1 +(\lambda_u +\lambda_c) (a_4+a_{10}) \right \}\right ] .\\
{\cal A} ( B^- \to K^- \rho^0)  &=&  \frac{G_F}{2} m_B^2 \left[
f_{K} A_0^{B\to \rho } (m_{K}^2)
\left \{ \lambda_u  a_1 +(\lambda_u +\lambda_c)[a_4+a_{10}
-r^{K}_{\chi}(a_6+a_8)]\right \}\right. \nonumber \\
&&+  \left .  f_\rho F_1^{B\to K }(m_\rho^2)
\left \{ \lambda_u a_2 +(\lambda_u +\lambda_c)\times\frac{3}{2}(a_9+a_{7}) \right \}\right].\\
{\cal A} ({\overline{B}}^{0} \to {\overline K}^0 \omega)   &=&
\frac{G_F}{2} m_B^2 
\left( f_{K} A_0^{B\to \omega } (m_{K^{0}}^2)(\lambda_u +\lambda_c)\left[a_4-\frac{1}{2}a_{10} -r^{K}_{\chi}(a_6-\frac{1}{2}a_8)\right] \right. \nonumber\\
&& +   \left . f_\omega F_1^{B\to K }(m_\omega^2)
\left \{ \lambda_u a_2 +(\lambda_u +\lambda_c)\left[2(a_3+a_5)+\frac{1}{2}(a_9+a_{7}) \right]
\right \} \right).~~~~~~~~~~~~~\\
{\cal A} ( B^- \to K^- \omega)  &=&  \frac{G_F}{2} m_B^2 \Bigl[
 f_{K} A_0^{B\to \omega } (m_{K}^2)\left \{ \lambda_u  a_1 +(\lambda_u +\lambda_c)(a_4+a_{10} -r^{K}_{\chi}(a_6+a_8))\right \}\Bigr.\nn\\
&&+ \Bigl.   f_\omega F_1^{B\to K }(m_\omega^2)
\left \{ \lambda_u a_2 +(\lambda_u +\lambda_c)\left(2(a_3+a_5)+\frac{1}{2}(a_9+a_{7})\right)
 \right \}\Bigr].~~~~~~~~~~~\\ 
{\cal A}( B^- \to K^{*-} \eta ^{(\prime)})  &=& \frac{G_F}{\sqrt{2}} m_B^2 \Bigl(
 f_{K^*} F_1^{B\to \eta  ^{(\prime)}} (m_{K}^2)\left \{ \lambda_u   a_1 
+(\lambda_u +\lambda_c)(a_4+a_{10} )\right \}
+ f_{\eta ^{(\prime)}}^u A_0^{B\to K^* }(m_{\eta ^{(\prime)}}^2) \label{eq:BtoK*}\\
&&\times
\left \{ \lambda_u  a_2 +\lambda_c  a_2 \frac{ f_{\eta ^{(\prime)}}^c}
{f_{\eta^{(\prime)} }^u}
+(\lambda_u +\lambda_c)\left[2(a_3-a_5)+\frac{1}{2}(a_9-a_{7})+r^{\eta(\prime)}_{\chi}(a_6-\frac{1}{2}a_{8})\right.\right.\nonumber\\
&&\left.\left.\left.
\hspace{-1.cm}+(a_3-a_5+a_9-a_7)\frac{f_{\eta^{(\prime)}}^c}{f_{\eta^{(\prime)}}^u}+
 \left(a_3-a_5-\frac{1}{2}(a_9-a_{7})+a_4-\frac{1}{2}a_{10}-r^{\eta(\prime)}_{\chi}
(a_6-\frac{1}{2}a_{8})\right)
\frac{f_{\eta ^{(\prime)}}^s}
{f_{\eta ^{(\prime)}}^u}\right]
 \right \}\right ). \nonumber\\
{\cal A}({\overline{B}}^{0} \to {\overline K}^{*0} \eta ^{(\prime)})  &=&  \frac{G_F}{\sqrt{2}} m_B^2 \left( 
 f_{K^*} F_1^{B\to \eta  ^{(\prime)}} (m_{K}^2)
(\lambda_u +\lambda_c)\left[a_4-\frac{1}{2} a_{10} \right] \right .
+ f_{\eta ^{(\prime)}}^u A_0^{B\to K^* }(m_{\eta
 ^{(\prime)}}^2) \\
&&\times \left \{ \lambda_u a_2
+\lambda_c  a_2 \frac{ f_{\eta ^{(\prime)}}^c} {f_{\eta^{(\prime)} }^u}
 + (\lambda_u +\lambda_c)\left[2(a_3-a_5)+\frac{1}{2}(a_9-a_{7})+r^{\eta(\prime)}_{\chi}
(a_6-\frac{1}{2}a_{8})+\right.\right.\nonumber \\
&& \left.\left.\left.
\hspace{-1.cm}+(a_3-a_5+a_9-a_7)\frac{f_{\eta^{(\prime)}}^c}
{f_{\eta^{(\prime)}}^u}
+ \left(a_3-a_5-\frac{1}{2}(a_9-a_{7})+a_4-\frac{1}{2}a_{10}-r^{\eta(\prime)}_{\chi}
(a_6-\frac{1}{2}a_{8})\right)\frac{f_{\eta ^{(\prime)}}^s}
{f_{\eta ^{(\prime)}}^u}\right]
 \right \} \right) . \nonumber
\end{eqnarray}

with $r^{\eta(\prime)}_{\chi}=\frac{2m_{\eta ^{(\prime)} }^2}{(m_b+m_s)(m_s+m_s)}$.
 \newpage \pagestyle{plain}
{\underline{\bf (3) Pure penguin processes:}}
\begin{eqnarray}
{\cal A} ( B^- \to \pi^- {\overline K}^{*0}) &=&\frac{G_F}{\sqrt{2}}
m_B^2 f_{K^*} F_1^{B\to \pi } (m_{K^{*}}^2) (\lambda_u +\lambda_c)  \left [
a_4-\frac{1}{2}a_{10} \right ].\\
  {\cal A} ( B^- \to \rho^- {\overline K}^0) &=&\frac{G_F}{\sqrt{2}}
m_B^2 f_{K} A_0^{B\to \rho } (m_{K^{0}}^2)  (\lambda_u +\lambda_c)
 \left [ a_4-\frac{1}{2}a_{10}-r^{K}_{\chi}(a_6-\frac{1}{2}a_8) \right ].\\
  {\cal A} (B^- \to K^-  K^{*0}) 
 &=&\frac{G_F}{\sqrt{2}}
m_B^2 f_{K^*} F_1^{B\to K } (m_{K^{*}}^2) (\lambda_u' +\lambda_c') 
\left [a_4-\frac{1}{2}a_{10} \right ].\\
  {\cal A} ( B^- \to K^{*-}  K^0)
&=&\frac{G_F}{\sqrt{2}} m_B^2
f_{K} A_0^{B\to K^* } (m_{K^{0}}^2) (\lambda_u' +\lambda_c')
\left [ a_4-\frac{1}{2}a_{10}-r^{K}_{\chi}(a_6-\frac{1}{2}a_8) \right ].\\
{\cal A} ( B^{-} \to \pi^- \phi) 
 &=&-\frac{G_F}{2} m_B^2 f_\phi F_1^{B\to \pi }(m_\phi^2) (\lambda_u' +\lambda_c')  
\left[ a_3+a_5-\frac{1}{2}(a_7+a_9)\right ].\\ 
{\cal A} ( B^- \to K^- \phi) &=& {\cal A} ({\overline{B}}^{0} \to {\overline K}^0 \phi) \nn\\
~~~~~~~~~~~&=& {G_F \over \sqrt{2}} m_B^2 
f_\phi F_1^{B\to K}(m_\phi^2) (\lambda_u +\lambda_c) 
\left[ a_3+a_4+a_5-\frac{1}{2}(a_7+a_9+a_{10})\right]. 
\end{eqnarray}


\vspace*{-0.75cm}
\section{The annihilation amplitudes for $B{\to}PV$}
\label{sec:app1}\hspace*{\parindent}
We give in this section the following annihilation amplitudes for
$B{\to}PV$ already given in ref.~\cite{Du:2002up} but with different
 notations:\\
{\underline{\bf (1) $b\to d$  processes:}}
 \begin{eqnarray}
{\cal A}^{a}({\overline{B}}^{0}{\to}{\pi}^{-}{\rho}^{+})
&=& \frac{G_{F}}{\sqrt{2}} f_{B} f_{\pi} f_{\rho}
  \bigg\{ \lambda_u' b_{1}({\rho}^{+},{\pi}^{-})
+ (\lambda_u' +  \lambda_c' )
  \Big[ b_{3}({\pi}^{-},{\rho}^{+})
      + b_{4}({\rho}^{+},{\pi}^{-})
      + b_{4}({\pi}^{-},{\rho}^{+})\nn\\
&&- \frac{1}{2}b_{3}^{ew}({\pi}^{-},{\rho}^{+})
      + b_{4}^{ew}({\rho}^{+},{\pi}^{-})
      - \frac{1}{2}b_{4}^{ew}({\pi}^{-},{\rho}^{+})
  \Big] \bigg\}.
 \label{eq:appendix-3}\\
 {\cal A}^{a}({\overline{B}}^{0}{\to}{\pi}^{+}{\rho}^{-})
  &=& \frac{G_{F}}{\sqrt{2}} f_{B} f_{\pi} f_{\rho}
  \bigg\{ \lambda_u' b_{1}({\pi}^{+},{\rho}^{-})
+ ( \lambda_u' +  \lambda_c' )
  \Big[ b_{3}({\rho}^{-},{\pi}^{+})
      + b_{4}({\pi}^{+},{\rho}^{-})
      + b_{4}({\rho}^{-},{\pi}^{+}) \nonumber \\
  & & - \frac{1}{2}b_{3}^{ew}({\rho}^{-},{\pi}^{+})
      + b_{4}^{ew}({\pi}^{+},{\rho}^{-})
      - \frac{1}{2}b_{4}^{ew}({\rho}^{-},{\pi}^{+})
  \Big] \bigg\}.
 \label{eq:appendix-4}\\
  {\cal A}^{a}({\overline{B}}^{0}{\to}{\pi}^{0}{\rho}^{0})
   &=& \frac{G_{F}}{2 \sqrt{2}} f_{B} f_{\pi} f_{\rho}
  \bigg\{ \lambda_u' 
  \Big[ b_{1}({\rho}^{0},{\pi}^{0})
      + b_{1}({\pi}^{0},{\rho}^{0}) \Big]
       \nn\\
   & & + (  \lambda_u'  +  \lambda_c' )
  \Big[ b_{3}({\rho}^{0},{\pi}^{0}) + b_{3}({\pi}^{0},{\rho}^{0})
+ 2b_{4}({\pi}^{0},{\rho}^{0}) + 2b_{4}({\rho}^{0},{\pi}^{0}) \nonumber \\
&&+\frac{1}{2}
\Big(-b_{3}^{ew}({\rho}^{0},{\pi}^{0})- b_{3}^{ew}({\pi}^{0},{\rho}^{0})
+ b_{4}^{ew}({\pi}^{0},{\rho}^{0})+ b_{4}^{ew}({\rho}^{0},{\pi}^{0})\Big)
  \Big] \bigg\}.\label{eq:appendix-12} \\
  {\cal A}^{a}(B^{-}{\to}{\pi}^{-}{\rho}^{0})
   &=& \frac{G_{F}}{2} f_{B} f_{\pi} f_{\rho}
  \bigg\{   \lambda_u' 
  \Big[ b_{2}({\pi}^{-},{\rho}^{0}) - b_{2}({\rho}^{0},{\pi}^{-}) \Big]
        \nn\\
   & & + ( \lambda_u'  +  \lambda_c')
  \Big[ b_{3}({\pi}^{-},{\rho}^{0}) - b_{3}({\rho}^{0},{\pi}^{-})
  + b_{3}^{ew}({\pi}^{-},{\rho}^{0})- b_{3}^{ew}({\rho}^{0},{\pi}^{-})
  \Big] \bigg\}.
 \label{eq:appendix-21}\\
  {\cal A}^{a}(B^{-}{\to}{\pi}^{0}{\rho}^{-})
   &=& \frac{G_{F}}{2} f_{B} f_{\pi} f_{\rho}
  \bigg\{ \lambda_u'
  \Big[ b_{2}({\rho}^{-},{\pi}^{0})
      - b_{2}({\pi}^{0},{\rho}^{-}) \Big]
       \nn\\
   & & + ( \lambda_u'  +  \lambda_c')
  \Big[ b_{3}({\rho}^{-},{\pi}^{0}) - b_{3}({\pi}^{0},{\rho}^{-})
 + b_{3}^{ew}({\rho}^{-},{\pi}^{0})- b_{3}^{ew}({\pi}^{0},{\rho}^{-})
  \Big] \bigg\}.
 \label{eq:appendix-250}\\
  {\cal A}^{a}(B^{-}{\to}{\pi}^{-}{\omega})
   &=& \frac{G_{F}}{2} f_{B} f_{\pi} f_{\omega}
  \bigg\{ \lambda_u'
  \Big[ b_{2}({\pi}^{-},{\omega}) + b_{2}({\omega},{\pi}^{-}) \Big]
        \nn\\
   & & + ( \lambda_u'  +  \lambda_c')
  \Big[ b_{3}({\pi}^{-},{\omega}) + b_{3}({\omega},{\pi}^{-})
+ b_{3}^{ew}({\pi}^{-},{\omega})+ b_{3}^{ew}({\omega},{\pi}^{-})
  \Big] \bigg\}. \label{eq:appendix-22}
 \end{eqnarray}

 \newpage \pagestyle{plain}
{\underline{\bf (2) $b\to s$  processes:}}
\begin{eqnarray}
 {\cal A}^{a}({\overline{B}}^{0}{\to}{\pi}^{+}K^{{\ast}-})
  &=& \frac{G_{F}}{\sqrt{2}} f_{B} f_{\pi} f_{K^{\ast}}
  \bigg\{ ( \lambda_u  +  \lambda_c)
  \Big[ b_{3}(K^{{\ast}-},{\pi}^{+})
- \frac{1}{2}b_{3}^{ew}(K^{{\ast}-},{\pi}^{+})
  \Big] \bigg\}.~~~\label{eq:appendix-5}\\
 {\cal A}^{a}({\overline{B}}^{0}{\to}K^{-}{\rho}^{+})
  &=& \frac{G_{F}}{\sqrt{2}} f_{B} f_{K} f_{\rho}
  \bigg\{  ( \lambda_u  +  \lambda_c)
  \Big[ b_{3}(K^{-},{\rho}^{+})
- \frac{1}{2}b_{3}^{ew}(K^{-},{\rho}^{+})
  \Big] \bigg\}.\label{eq:appendix-7}\\
  {\cal A}^{a}({\overline{B}}^{0}{\to}{\overline{K}}^{0}{\rho}^{0})
   &=& - \frac{G_{F}}{2} f_{B} f_{K} f_{\rho}
  \bigg\{  ( \lambda_u  +  \lambda_c)
  \Big[ b_{3}({\overline{K}}^{0},{\rho}^{0})
      - \frac{1}{2}b_{3}^{ew}({\overline{K}}^{0},{\rho}^{0})
  \Big] \bigg\}.
 \label{eq:appendix-10}\\
  {\cal A}^{a}({\overline{B}}^{0}{\to}{\overline{K}}^{0}{\omega})
   &=& \frac{G_{F}}{2} f_{B} f_{K} f_{\omega}
  \bigg\{ ( \lambda_u  +  \lambda_c)
  \Big[ b_{3}({\overline{K}}^{0},{\omega})
      - \frac{1}{2}b_{3}^{ew}({\overline{K}}^{0},{\omega})
  \Big] \bigg\}.
 \label{eq:appendix-11}\\
  {\cal A}^{a}(B^{-}{\to}K^{-}{\omega})
   &=& \frac{G_{F}}{2} f_{B} f_{K} f_{\omega}   
  \bigg\{  \lambda_u b_{2}(K^{-},{\omega})+ ( \lambda_u  +  \lambda_c)
  \Big[ b_{3}(K^{-},{\omega}) + b_{3}^{ew}(K^{-},{\omega})
  \Big] \bigg\}.~~~~~~\label{eq:appendix-31}\\
  {\cal A}^{a}(B^{-}{\to}{\pi}^{0}K^{{\ast}-})
   &=& \frac{G_{F}}{2} f_{B} f_{\pi} f_{K^{\ast}}
  \bigg\{  \lambda_u b_{2}(K^{{\ast}-},{\pi}^{0})+ ( \lambda_u  +  \lambda_c)
  \Big[ b_{3}(K^{{\ast}-},{\pi}^{0})
      + b_{3}^{ew}(K^{{\ast}-},{\pi}^{0})
  \Big] \bigg\}.~~~~~~\label{eq:appendix-26}\\
  {\cal A}^{a}(B^{-}{\to}K^{-}{\rho}^{0})
   &=& \frac{G_{F}}{2} f_{B} f_{K} f_{\rho}   
  \bigg\{ \lambda_u b_{2}(K^{-},{\rho}^{0})+ ( \lambda_u  +  \lambda_c)
  \Big[ b_{3}(K^{-},{\rho}^{0}) + b_{3}^{ew}(K^{-},{\rho}^{0})
  \Big] \bigg\}.~~~~~~\label{eq:appendix-32}\\
{\cal A}^{a}({\overline{B}}^{0}{\to}{\eta}^{({\prime})}{\overline{K}}^{{\ast}0})
   &=& \frac{G_{F}}{\sqrt{2}} f_{B} f_{{\eta}^{({\prime})}}^{u} f_{K^{\ast}}
  \bigg\{ ( \lambda_u  +  \lambda_c)
  \Big[ b_{3}({\overline{K}}^{{\ast}0},{\eta}^{({\prime})})
        \label{eq:appendix-19}\\
   & & - \frac{1}{2}b_{3}^{ew}({\overline{K}}^{{\ast}0},{\eta}^{({\prime})})
       + \frac{f_{{\eta}^{({\prime})}}^{s}}{f_{{\eta}^{({\prime})}}^{u}}
  \Big( b_{3}({\eta}^{({\prime})},{\overline{K}}^{{\ast}0})
      - \frac{1}{2}b_{3}^{ew}({\eta}^{({\prime})},{\overline{K}}^{{\ast}0}) \Big)
  \Big] \bigg\}.
 \nn \\
  {\cal A}^{a}(B^{-}{\to}{\eta}^{({\prime})}K^{{\ast}-})
   &=& \frac{G_{F}}{\sqrt{2}} f_{B} f_{{\eta}^{({\prime})}}^{u} f_{K^{\ast}}
  \bigg\{ \lambda_u  \Big[ b_{2}(K^{{\ast}-},{\eta}^{({\prime})})
     + \frac{f_{{\eta}^{({\prime})}}^{s}}{f_{{\eta}^{({\prime})}}^{u}}
       b_{2}({\eta}^{({\prime})},K^{{\ast}-}) \Big]
       \label{eq:appendix-28} \\
   & & + ( \lambda_u  +  \lambda_c)
  \Big[ b_{3}(K^{{\ast}-},{\eta}^{({\prime})})
      + b_{3}^{ew}(K^{{\ast}-},{\eta}^{({\prime})})
+ \frac{f_{{\eta}^{({\prime})}}^{s}}{f_{{\eta}^{({\prime})}}^{u}}
  \Big( b_{3}({\eta}^{({\prime})},K^{{\ast}-})
      + b_{3}^{ew}({\eta}^{({\prime})},K^{{\ast}-}) \Big)
  \Big] \bigg\}.\nn
 \end{eqnarray}

{\underline{\bf (3) Pure penguin processes:}}
\begin{eqnarray}
  {\cal A}^{a}(B^{-}{\to}{\pi}^{-}{\overline{K}}^{{\ast}0})
   &=& \frac{G_{F}}{\sqrt{2}} f_{B} f_{\pi} f_{K^{\ast}}
  \bigg\{  \lambda_u b_{2}({\overline{K}}^{{\ast}0},{\pi}^{-})
+ (\lambda_u  +  \lambda_c)
  \Big[ b_{3}({\overline{K}}^{{\ast}0},{\pi}^{-})
      + b_{3}^{ew}({\overline{K}}^{{\ast}0},{\pi}^{-})
  \Big] \bigg\}.\label{eq:appendix-24}\\
  {\cal A}^{a}(B^{-}{\to}{\overline{K}}^{0}{\rho}^{-})
   &=& \frac{G_{F}}{\sqrt{2}} f_{B} f_{K} f_{\rho}
  \bigg\{  \lambda_u b_{2}({\overline{K}}^{0},{\rho}^{-})
+ (\lambda_u  +  \lambda_c)
  \Big[ b_{3}({\overline{K}}^{0},{\rho}^{-})
      + b_{3}^{ew}({\overline{K}}^{0},{\rho}^{-})
\Big] \bigg\}.\label{eq:appendix-25}\\
  {\cal A}^{a}(B^{-}{\to}K^{-}K^{{\ast}0})
   &=& \frac{G_{F}}{\sqrt{2}} f_{B} f_{K} f_{K^{\ast}}
  \bigg\{ \lambda_u' b_{2}(K^{{\ast}0},K^{-})
  + (\lambda_u'  +  \lambda_c')
  \Big[ b_{3}(K^{{\ast}0},K^{-}) + b_{3}^{ew}(K^{{\ast}0},K^{-})
  \Big] \bigg\}.~~~~~~\label{eq:appendix-29}\\
  {\cal A}^{a}(B^{-}{\to}K^{0}K^{{\ast}-})
&=& \frac{G_{F}}{\sqrt{2}} f_{B} f_{K} f_{K^{\ast}}
  \bigg\{ \lambda_u' b_{2}(K^{0},K^{{\ast}-})
  + (\lambda_u'  +  \lambda_c')
  \Big[ b_{3}(K^{0},K^{{\ast}-}) + b_{3}^{ew}(K^{0},K^{{\ast}-})
  \Big] \bigg\}.~~~~~~\label{eq:appendix-20}\\
  {\cal A}^{a}(B^{-}{\to}{\pi}^{-}{\phi})&=& {\cal A}^{a}({\overline{B}}^{0}{\to}{\pi}^{0}{\phi})=0.
 \label{eq:appendix-23}\\
  {\cal A}^{a}(B^{-}{\to}K^{-}{\phi})
   &=& \frac{G_{F}}{\sqrt{2}} f_{B} f_{K} f_{\phi}
  \bigg\{ \lambda_u b_{2}({\phi},K^{-})
 + (\lambda_u  +  \lambda_c)
  \Big[ b_{3}({\phi},K^{-}) + b_{3}^{ew}({\phi},K^{-})
  \Big] \bigg\}.\label{eq:appendix-30}\\
  {\cal A}^{a}({\overline{B}}^{0}{\to}{\overline{K}}^{0}{\phi})
   &=& \frac{G_{F}}{\sqrt{2}} f_{B} f_{K} f_{\phi}
  \bigg\{ (\lambda_u  +  \lambda_c)
  \Big[ b_{3}({\phi},{\overline{K}}^{0})
      - \frac{1}{2}b_{3}^{ew}({\phi},{\overline{K}}^{0})
  \Big] \bigg\}.
 \label{eq:appendix-9}
 \end{eqnarray}
\end{appendix}


 
\end{document}